\documentclass[11pt]{article}
\usepackage{epsfig}
\usepackage{amssymb}
\usepackage[]{times}
\usepackage{xcolor}

\makeatletter \@addtoreset{equation}{section}

\makeatother
\newcommand{\be}{\begin{equation}}
\newcommand{\ee}{\end{equation}}
\newcommand{\bea}{\begin{eqnarray}}
\newcommand{\eea}{\end{eqnarray}}

\newtheorem{conjecture}{Conjecture}

\def\1#1{^{(#1)}}

\begin{document}
%%%%%%%%%%%%%%%%%%%%%%%%%%%%%%%%%%%%%%%%%%%%%%%%%%%%
\title{Isotropic conductivity of two-dimensional \\three- and four-phase symmetric composites: \\duality and universal bounds}
\author{Leonid G. Fel\\ \\
Department of Civil Engineering, Technion - Israel Institute of Technology,\\
Haifa, 32000, Israel\\ 
%{\em e-mail: lfel@technion.ac.il}
}
\date{}
%%%%%%%%%%%%%%%%%%%%%%%%%%%%%%%%%%%%%%%%%%%%%%%%%%%%
\maketitle
%%%%%%%%%%%%%%%%%%%%%%%%%%%%%%%%%%%%%%%%%%%%%%%%%%%%%%%%%%%%%%%%%%%%%%
\begin{abstract}
We consider the problem of isotropic effective conductivity $\sigma
_e(\sigma_1,\ldots,\sigma_n)$ in two-dimensional three- and four-phase symmetric composites with a partial isotropic conductivity $\sigma_j$ of the $j$-th phase. The upper $\Omega(\sigma_1,\ldots,
\sigma_n)$ and lower $\omega(\sigma_1,\ldots,\sigma_n)$, $n=3,4$, bounds for effective conductivity, found by the algebraic approach in \cite{fel22}, are universal (independent of the composite microstructure) and possess all algebraic properties of $\sigma_e(
\sigma_1,\ldots,\sigma_n)$ that follow from physics: first-order homogeneity, full permutation invariance, Keller's self-duality, positivity, and monotony. The bounds are compatible with the trivial solution $\sigma_e(\sigma,\ldots,\sigma)=\sigma$ and satisfy Dykhne's ansatz. Their comparison with previously known numerical calculations, asymptotic analysis, and exact results for isotropic effective conductivity $\sigma_e(\sigma_1,\ldots,\sigma_n
)$ of two-dimensional three- and four-phase composites showed complete agreement. The bounds $\Omega(\sigma_1,\ldots,\sigma_n)$ and $\omega(\sigma_1,\ldots,\sigma_n)$ in both cases $n=3,4$ are stronger than the currently known variational bounds.
{\bf Pacs:} 72.15.Eb, 72.80.Tm, 61.50.Ah
\end{abstract}
%%%%%%%%%%%%%%%%%%%%%%%%%%%%%%%%%%%%%%%%%%%%%%%%%%%%%%%%%%%
\noindent
%\hfill
%\widetext
%\narrowtext
%%%%%%%%%%%%%%%%%%%%%%%%%%%
\section{Introduction}\label{part1}
%%%%%%%%%%%%%%%%%%%%%%%%%%%%%%%%%%%%%%%%%%%%%%%%%%%%%%%%%%%
A dual symmetry of isotropic effective conductivity $\sigma_e(
\sigma_1,\sigma_2)$ in two-dimensional (2D) two-phase composites, discovered by Keller \cite{kel64} and Dykhne \cite{dyk70},
\bea
\sigma_e(\sigma_1,\sigma_2)\cdot\sigma_e(\sigma_1^{-1},
\sigma_2^{-1})=1,\label{a1}
\eea
has shown a way to find $\sigma_e(\sigma_1,\sigma_2)$ even without solving Maxwell's equations. The power of this method was demonstrated by the fact that the universal isotropic solution 
$\sqrt{\sigma_1\sigma_2}$ for a two-phase composite with equal volume fractions $p_1\!=\!p_2\!=\!1/2$ of constituents is valid for every regular structure (translationally symmetric in the plane, often called {\em the two-color checkerboard}) with square  \cite{ber85,eme89,eob89} and proper triangle \cite{obn99} unit cells. The universal solution is valid also for random structures \cite{dyk70,men75,koz89,hel11}. In 2D two-phase composites, a dual symmetry has also been applied to coupled divergence-free vector fields \cite{bal82,fel02a}, for a regular composite with rectangular \cite{obn99,obn96} and rhombic \cite{fel03,ovc04} unit cells, where instead of equation (\ref{a1}), its analogues for two eigenvalues of the anisotropic tensor $\sigma_e^{ij}(\sigma_1,\sigma_2)$ are valid \cite{kel64,fel02b}. 

Further progress in the isotropic effective conductivity (EC) problem in 2D composites was achieved in \cite{fel00} by expanding the duality relation in the 2D $n$-phase ($n>2$) composite
\bea
\sigma_e(\sigma_1,\ldots,\sigma_n)\cdot\sigma_e(\sigma_1^{-1},\ldots,\sigma_n^{-1})=1.\label{a2}
\eea
The functional equation (\ref{a2}) does not provide a universal isotropic solution for various regular and random composites (see four three-color checkerboards with different unit cells \cite{fel00}) and leaves open the question of explicit solutions for the 2D $n$-phase composites with a given shape of unit cells. Meanwhile, such solutions (the two eigenvalues of $\sigma_e^{ij}(\sigma_1,\ldots,\sigma_n)$, $n=3,4$) were found in three-phase composites with diamond \cite{cra04} and in four-phase composites with square \cite{mor85,mil01,cra01,cra06} unit cells. These works were based on analytical solutions of the Riemann-Hilbert boundary value problem with the corresponding conformal mapping in the complex plane. However, analytical solutions to the EC problem with a large number of phases and a highly complex unit cell encounter enormous technical difficulties.

The absence of universal solutions to the EC problem in 2D $n$-phase composites has one more substantial consequence, specifically, the absence of uniqueness of the effective conductivity $\sigma_e(\sigma_1,p_1;\sigma_2,p_2)$, where $p_1\!\ne\!p_2$, even in the 2D two-phase compound with a different distribution of the constituents. Indeed, equating two partial conductivities $\sigma_2\!=\!\sigma_3$ in four different three-color checkerboards \cite{fel00}, we arrive at $p_1=1/3,p_2=2/3$ and find four different dependences for $\sigma_e(\sigma_1/\sigma_2)$.

The analytical approach, as well as numerical calculations \cite{pas94,hel98,lam24} and the asymptotic expansion approach \cite{koz89,kel87} or the network analogue \cite{luc91,felk02,
wan07} all suffer from non-universality. Therefore, they are usually accompanied by a variational bound approach and methods of compensated compactness. The best known isotropic bounds were obtained by Wiener \cite{wie12}, Hashin-Shtrikman \cite{has62}, Nesi \cite{nes91,nes95}, and Cherkaev \cite{che09,che12,che24}, where the number of phases can be arbitrary. 
The theory was also developed for anisotropic composites built of isotropic \cite{tar85,lur85} or anisotropic phases \cite{mik88}. 
Although isotropic bounds of effective conductivity $\sigma_e$ are universal, some of them still lack a dual symmetry (\ref{a2}). 
In the next Section \ref{part12}, we discuss an algebraic approach that involves such symmetry.
%%%%%%%%%%%%%%%%%%%%%%%%%%%%%%%%%%%%%%
%%%%%%%%%%%%%%%%%%%%%%%%%%%%%%%%%%%%%%
\subsection{Algebraic approach in the EC problem for 2D three-phase composites}\label{part12}
%%%%%%%%%%%%%%%%%%%%%%%%%%%%%%%%%%%%%%
%%%%%%%%%%%%%%%%%%%%%%%%%%%%%%%%%%%%%%
To study the EC problem in 2D three-phase regular composites, a cubic equation for $\sigma_e$ with coefficients, built on three symmetric polynomials $I_j$ and one free parameter $A\ge 0$, was proposed in \cite{fel00} 
\bea
\sigma_e^3+AI_1\sigma_e^2-AI_2\sigma_e-I_3=0,\qquad I_1=\sum_{j=1}^
3\sigma_j,\quad I_2=\!\sum_{k>j=1}^3\!\sigma_j\sigma_k,\quad I_3=
\prod_{j=1}^3\sigma_j.\label{a3}
\eea
The cubic equation (\ref{a3}) implements the basic algebraic properties of the function $\sigma_e(\sigma_1,\sigma_2,\sigma_3)$ 
\footnote{A proof of monotony of solution $\sigma_e(\sigma_1,\sigma_2,\sigma_3)$ of equation (\ref{a3}), omitted in \cite{fel00}, may be obtained by opposite assumption: let one of derivatives be negative, e.g. $\partial\sigma_e/\partial\sigma_1<0
$. Since in the region $\sigma_i\geq 0$, there always exist lines 
where $\partial\sigma_e/\partial\sigma_1>0$, e.g. when $\sigma_1=
\sigma_2=\sigma_3$, then necessarily there must exist a point 
$(\sigma_1^{*},\sigma_2^{*},\sigma_3^{*})$, where $\partial\sigma_e
/\partial\sigma_1 =0$. Differentiating equation (\ref{a3}) by 
$\sigma_1$ and combining the result once more with (\ref{a3}), we get: $\sigma_{e}^3+A\sigma_{e}(\sigma_2^{*2}+\sigma_3^{*2}+\sigma_2
^{*}\sigma_3^{*})+(\sigma_2^{*}+\sigma_3^{*})\sigma_2^{*}\sigma_3^
{*}=0$. Since $\sigma_{e}\geq 0, A\geq 0$, the last equation has not positive solutions $(\sigma_1^{*},\sigma_2^{*},\sigma_3^{*})$, and our assumption is wrong.} :
\bea
&&\hspace{-.6cm}
a)\;\;\mbox{homogeneity of the 1st order},\quad b)\;\;S_3- \mbox{color permutation invariance},\quad 
c)\;\;\mbox{self-dual symmetry},\nonumber\\
&&\hspace{-.6cm}
d)\;\;\mbox{monotony w.r.t. any variable $\sigma_j$ of}\;\;\sigma_e(\sigma_1,\sigma_2,\sigma_3),\quad 
e)\;\;\mbox{compatibility}\;\;\sigma_e(\sigma,\sigma,\sigma)=\sigma,\nonumber\\
&&\hspace{-.6cm}
f)\;\;\sigma_e\;\;\mbox{satisfies Dykhne's ansatz\; \cite{dyk70}}:
\quad\mbox{if}\;\;\sigma_1\cdot\sigma_2=\sigma_3^2\quad\mbox{then}
\quad\sigma_e(\sigma_1,\sigma_2,\sigma_3)=\sigma_3,\label{a4}
\eea
and coincides with Bruggeman's equation \cite{bru35} if $A\!=
\!1/3$ in the effective medium approximation EMA.
%%%%%%%%%%%%%%%%%%%%%%%%%%%%%%%%%%%%%%%%%%%
%%%%%%%%%%%%%%%%%%%%%%%%%%%%%%%%%%%%%%%%%%%
\begin{figure}[h!]\begin{center}
\psfig{figure=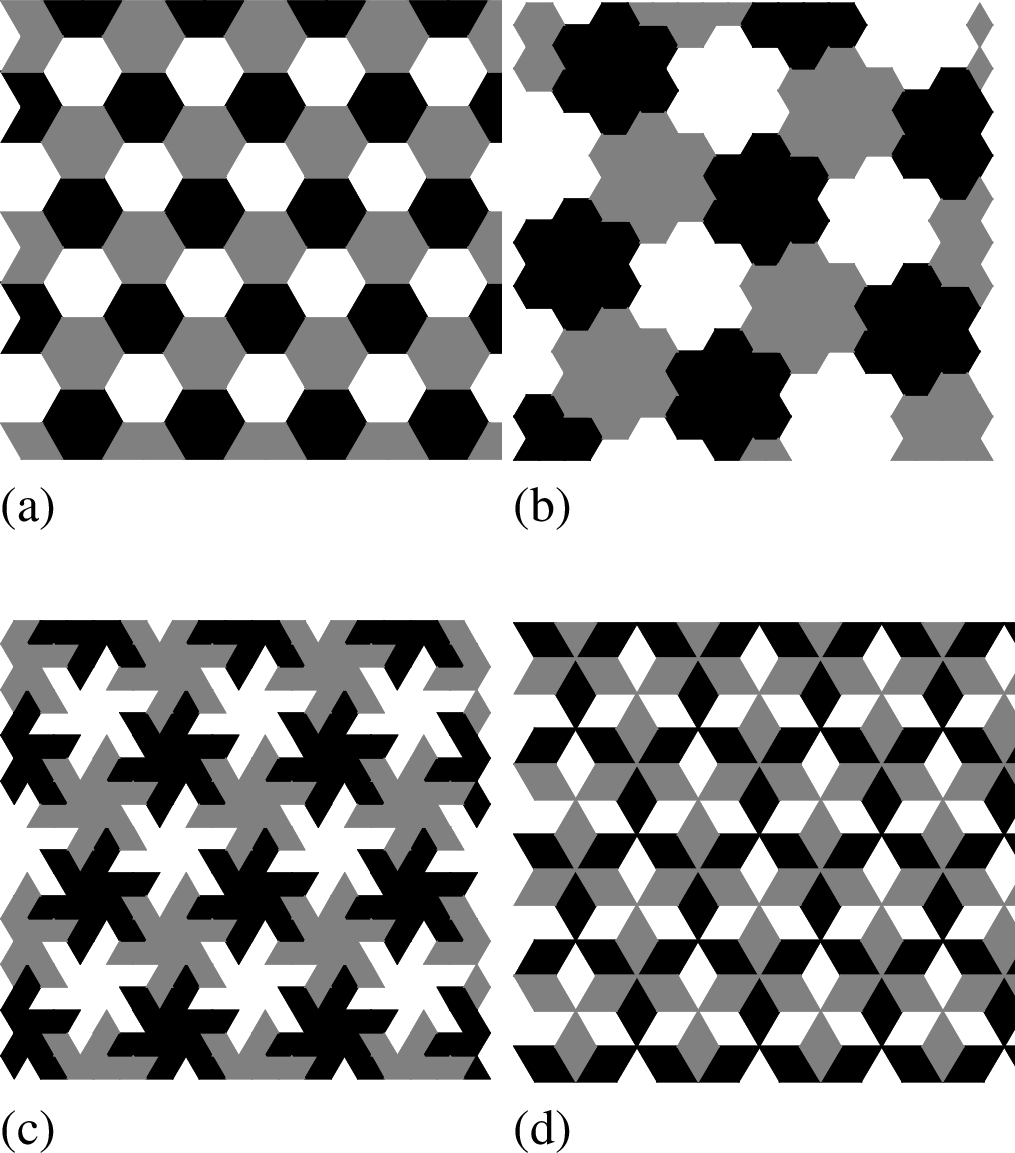,height=9cm}
\end{center}
\caption{Three phase 2D composites \cite{fel00}: a) {\sf He} - $A_{He}=11.37$, $\epsilon_{max}(He)\!=\!0.00026$; b) {\sf Fl} - $A_{Fl}=3.76$, $\epsilon_{max}(Fl)\!=\!0.001$; c) {\sf Co} - $A_{Co}=0.305$, $\epsilon_{max}(Co)\!=\!0.022$; d) {\sf Rh} - $A_{Rh}\!=\!0$, $\epsilon_{max}(Rh)\!=\!0.065$.}\label{fig1}
\end{figure}
%%%%%%%%%%%%%%%%%%%%%%%%%%%%%%%%%%%%%%%%%%%
%%%%%%%%%%%%%%%%%%%%%%%%%%%%%%%%%%%%%%%%%%%

Support for this algebraic approach was found in \cite{fel00}, which used numerical calculations for four different infinite 2D three-phase composites with the 3-fold rotation axis and $S_3$-permutation invariance of the phases (see Figure \ref{fig1}). Equation (\ref{a3}) governs the EC problem in all four structures with very high precision $\epsilon_{max}\simeq 10^{-4}-10^{-2}$. Here, $\epsilon_{max}$ is the maximal relative deviation of the numerically calculated $\sigma_e$ from the solution of the cubic equation (\ref{a3}) with a certain free parameter $A$ that scans the entire range of $\sigma_j$: $0\le \sigma_1/\sigma_3\leq 1$, $\;0\le\sigma_2/\sigma_3\le 1$. For convenience, we assume $\sigma_1
\le\sigma_2\le\sigma_3$.

Despite such high accuracy, the paper \cite{fel00} did not claim equation (\ref{a3}) as an explicit and exact equation for the 2D three-phase EC problem. In fact, it was argued that equation (\ref{a3}) is just a minimal-order first algebraic approximation with one free parameter $A$, which gives an excellent agreement with calculations for various regular structures but lacks universality. Furthermore, this approximation is far from Bruggeman's equation \cite{bru35}. Beyond the scope of the paper \cite{fel00}, several questions remain open. 
%%%%%%%%%%%%%%%%%%%%%%%%%%%%%%%%%%%%%%%%%%%
%%%%%%%%%%%%%%%%%%%%%%%%%%%%%%%%%%%%%%%%%%%
\begin{enumerate}
\item Are there other algebraic equations of order $m>3$ with inherited basic properties (\ref{a4}) for the isotropic EC problem in 2D three-phase regular composites? If so, what is the underlying algebraic structure of these equations?
\item Are there lower $\omega(\sigma_1,\sigma_2,\sigma_3)$ and upper $\Omega(\sigma_1,\sigma_2,\sigma_3)$ universal bounds with inherited basic properties (\ref{a4}) for $\sigma_e(\sigma_1,\sigma_2,\sigma_3)$ in a 2D three-phase composite ?
\item Can such algebraic approach be extended to the case of a 2D $n$-phase composite with arbitrary number of phases ?
\end{enumerate}
%%%%%%%%%%%%%%%%%%%%%%%%%%%%%%%%%%%%%%%%%%%
%%%%%%%%%%%%%%%%%%%%%%%%%%%%%%%%%%%%%%%%%%%
These questions were addressed in a recent paper \cite{fel22}, in which a commutative monoid of self-dual polynomials ${\sf R}\left(^{\lambda,\;{\bf x^n}}_{m,\;S_n}\right)$ and  ${\sf S}\left(^
{\lambda,\;{\bf x^n}}_{m,\;S_n}\right)$ in $\lambda$, of degree $mn$, $m\in {\mathbb N}$, was constructed. Their coefficients are based on polynomial invariants $I_r({\bf x}^n)$ of the symmetric group $S_n$, acting on the Euclidean space ${\mathbb E}^n$, where 
${\bf x^n}=\{x_1,\ldots,x_n\}\in{\mathbb E}^n$. The real roots 
$\lambda\left({\bf x^n}\right)$ of the polynomials ${\sf S}\left(^{
\lambda,\;{\bf x^n}}_{m,\;S_n}\right)$ have many remarkable properties : first-order homogeneity, a self-dual symmetry under inversion of variables $x_i\to x_i^{-1}$ and function $\lambda\to
\lambda^{-1}$, monotony of $\lambda\left({\bf x^n}\right)$ with respect to every $x_i$ and others. In this regard, a cubic equation (\ref{a3}) represents the equation ${\sf S}\left(^{\lambda,\;{\bf 
x^3}}_{1,\;S_3}\right)=0$ with a minimal $m=1$, i.e., of degree three.

Polynomials ${\sf S}\left(^{\lambda,\;{\bf x^2}}_{m,\;S_2}\right)$ are always divisible by ${\sf S}\left(^{\lambda,\;{\bf x^2}}_{1,\;
S_2}\right)$, a fact that distinguishes the case $n=2$ from the others $n\ge 3$. On the other hand, by additionally requiring the existence of a unique positive solution of equation ${\sf S}\left(^
{\lambda,\;{\bf x^n}}_{m,\;S_n}\right)=0$, we were able
\cite{fel22} to find universal (for arbitrary $m$) bounds for the solution in the cases $n=3,4$. The problem of universal bounds for $n\ge 5$ remains open.

\begin{figure}[h]
\centerline{\psfig{figure=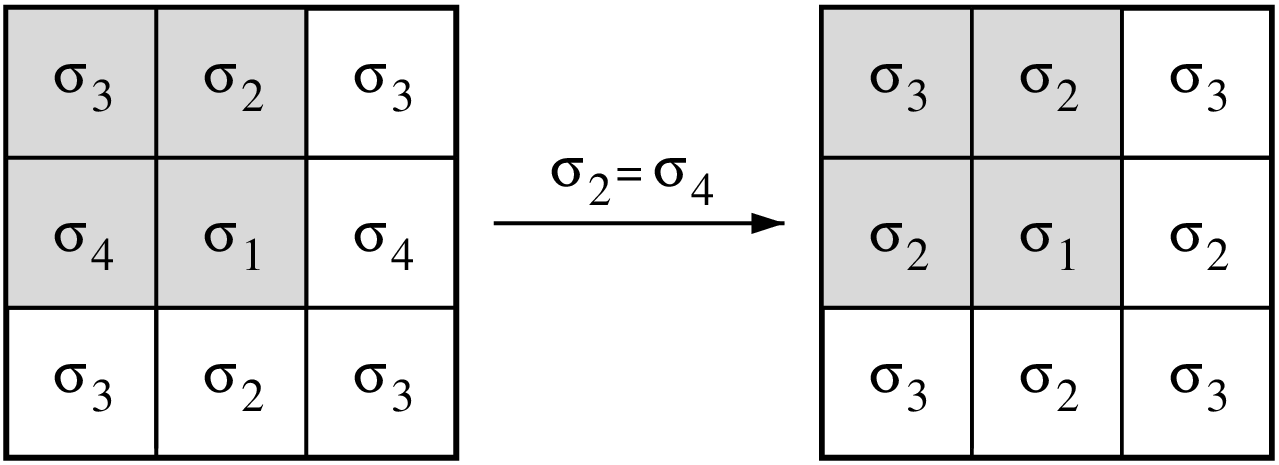,height=4cm}}
\caption{Four-color Mortola-Steff\'e checkerboard ({\sl left}) with 2-fold rotation axis and three-phase checkerboard ({\sl right}) with 4-fold rotation axis. The unit cells are dashed.} \label{fig2}
\end{figure}

In the present paper, we verify our approach by comparing the obtained bounds $\omega(\sigma_1,\ldots,\sigma_n)$ and $\Omega(
\sigma_1,\ldots,\sigma_n)$ for $n=3,4$, with the known variational, numerical, asymptotic, and exact results for the isotropic effective conductivity of 2D three- and four-phase composites. Such results mainly refer to the case $n=3$. The proof 
\cite{mil01,cra01} of Mortola-Steff\'e conjecture \cite{mor85} for the conductivity of four-color checkerboards with a square unit cell (see Figure \ref{fig2}) allows one to check the bounds also for $n=4$. 
%%%%%%%%%%%%%%%%%%%%%%%%%%%%%%%%%
%%%%%%%%%%%%%%%%%%%%%%%%%%%%%%%%%%%%%%%%%%%
%%%%%%%%%%%%%%%%%%%%%%%%%%%%%%%%%%%%
%%%%%%%%%%%%%%%%%%%%%%%%%%%%%%%%%%%%%%%%%%%

We show that the fundamental properties (\ref{a4}) provide a solid basis for achieving bounds that are significantly stronger than the currently known bounds.
%%%%%%%%%%%%%%%%%%%%%%%%%%%%%%%%%%%%%%%%%%%
%%%%%%%%%%%%%%%%%%%%%%%%%%%%%%%%%%%%%%%%%%%%%%%%%%%%
\section{Crystallography of color checkerboards}\label{part2}
%%%%%%%%%%%%%%%%%%%%%%%%%%%%%%%%%%%%%%%%%%%%%%%%%%%%
%%%%%%%%%%%%%%%%%%%%%%%%%%%%%%%%%%%%%%%%%%%%
The EC problem for multiphase 2D regular composites was treated \cite{fel00} using {\it n-color} plane groups. Different colors correspond to different phases. Following Shubnikov, we consider only {\it perfect coloring} of the checkerboard \cite{sen79}: the sectors of each color must form identical patterns, which are related to each other by a symmetry operation. Such a color tessellation is completely determined by the tiling and coloring rules: it is invariant under the action of the color plane group $[{\sf G}_1,{\sf P}_n,{\sf T}]$, which includes a 2D lattice point group ${\sf G}_1$, a color permutation group ${\sf P}_n$ and a translation group ${\sf T}$. Each color plane group has its origin in one of the 2D lattice plane groups $[{\sf G}_1,{\sf T}]$. The number $N_n$ of the {\it n-color} plane groups \cite{shw80} is a nonmonotone function of $n$: $N_1=17$, $N_2=46$, $N_3=23$, $N_4=96$, and $N_5=14$. The arrangement of color patterns within the unit cell varies for different color groups with the same $n$. 

In this paper, we are interested in the isotropic conductivity 
$\sigma_e$ of 2D $n$-phase composites with equal phase fractions, when each tile is presented with its isotropic $\sigma_j$. Following Curie's symmetry principle \cite{cir84}, we distinguish two closely related concepts: an isotropic 2D media with a point group $O(2)$ and an isotropic tensor in the media with a point group $G\subset O(2)$. According to Hermann's theorem \cite{her34} on tensors of rank $k$ in media with inner symmetry $G$ having a rotation axis ${\sf h}={\sf C}_r$, $\;r>k$, the presence of rotation axes ${\sf C}_r$, $r\geq 3$, makes the 2nd rank tensor $\sigma_e^{ij}$ isotropic. The highest order rotation axis in the 2D lattice is ${\sf h}_1={\sf C}_6$. By studying the tables of color plane groups \cite{shw80}, one can establish the highest-order rotation axis ${\sf h}_n$ allowed in $n$-color checkerboards: ${\sf h}_2={\sf C}_6,\;{\sf h}_3={\sf C}_3,\;{\sf h}_4={\sf C}_4,\;{\sf h}_5={\sf C}_2,\;{\sf h}_6={\sf C}_3$.
%%%%%%%%%%%%%%%%%%%%%%%%%%%%%%%%%%%%%%%%%%%
%%%%%%%%%%%%%%%%%%%%%%%%%%%%%%%%%%%%%%%%%%%
\begin{figure}[h]
\centerline{\psfig{figure=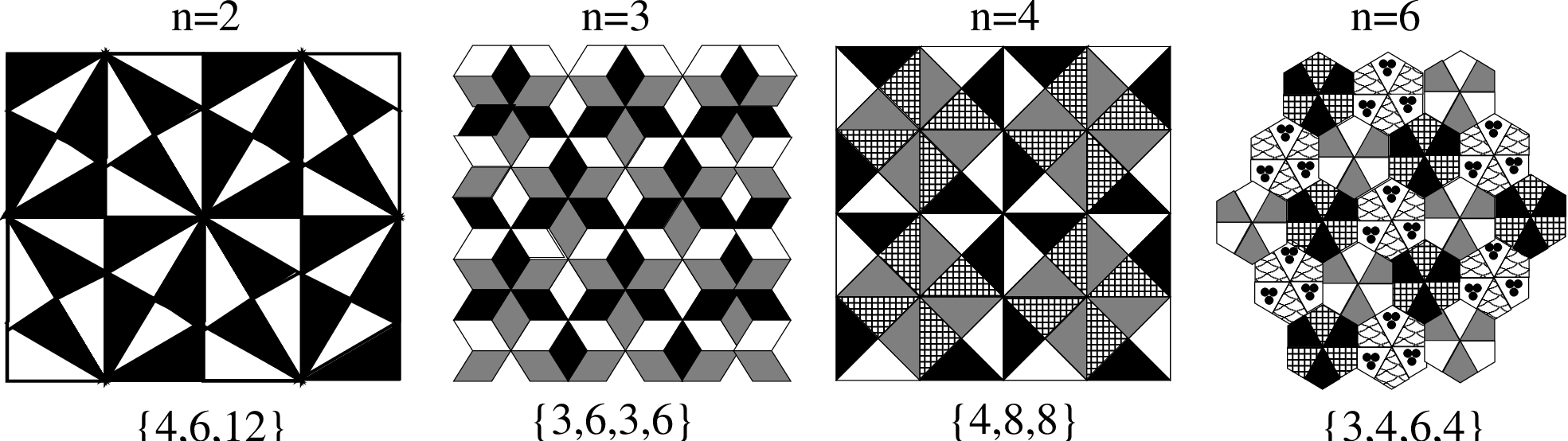,height=4cm}}
\caption{Four regular color tessellation of the plane: $\!\!$ $\{4,
6,12\}- {\sf h}_2={\sf C}_6,{\sf P}_2={\sf S_2};\;$ $\{3,6,3,6\} -
 {\sf h}_3={\sf C}_3,{\sf P}_3=S_3;\;$ $\{4,8,8\}- {\sf h}_4={\sf C}_4,{\sf P}_4={\sf D}_2;\;$ $\{3,4,6,4\}- {\sf h}_6={\sf C}_3,
{\sf P}_6={\sf D}_3$. They are associated with isotropic tensor of the 2nd rank and belong to the 11 topological types of isohedral face-to face plane tesselation \cite{eng86}. Their labels $\{v_1,\ldots, v_L\}$ denote the valences $v_j$ of unit $L$- gon at its $j$-th vertex.}\label{fig3}
\end{figure}
%%%%%%%%%%%%%%%%%%%%%%%%%%%%%%%%%%%%%%%%%%%
%%%%%%%%%%%%%%%%%%%%%%%%%%%%%%%%%%%%%%%%%%%

Another crystallographic restriction comes from the color permutation group ${\sf P}_n\subseteq S_n$. Let ${\sf G}_1$ be a point symmetry group of the 2D lattice. A checkerboard will have a permutational color point group $[{\sf G}_1,{\sf P}_n]$, if each of its tiles has a single color and if the tiles are distributed in such a way that an element of group ${\sf P}_n$, which permutes $n$ colors transitively, can be combined with an element of ${\sf G}_1$ to form a subgroup of the point group $[{\sf G}_1,{\sf P}_n]$, which restores the object to a condition indistinguishable from its original state. 

Let $H_1\subset {\sf G}_1$ be the subgroup keeping the first color fixed. For each subgroup $H_1$ of the finite index $\left|{\sf G}_1/H_1\right|$ there exists a subgroup $H\subseteq H_1$ which is a maximal normal subgroup of ${\sf G}_1$ contained in $H_1$, where $|H|$ denotes the order of $H$. It is known \cite{wan61} that ${\sf P}_n$ is isomorphic to the factor group ${\sf G}_1/H$ and therefore $|{\sf G}_1/H|=|{\sf P}_n|\leq |S_n|=n!$, where $|H|\ge 2$. Since the dihedral point symmetry group ${\sf G}_1={\sf D}_6$ in the plane has the highest order of $12$, we conclude that the regular $n$-color checkerboard of proper tiles, invariant under the full color permutation group $S_n$, can only be achieved for $n\leq 3$. 

The four-color checkerboard deserves special attention. On the one hand, according to the previous paragraph, there is no four-color tiling of the plane with color permutation group ${\sf P}_4=
S_4$. On the other hand, the existence of the Mortola-Steff\'e checkerboard (Figure \ref{fig2}) with dihedral color permutation group ${\sf P}_4={\sf D}_2$ and an anisotropic tensor $\sigma_{MS}^{ij}$ allows us to convert the latter into an isotropic one 
$\sigma_{MS}$ by equating two partial conductivities. It turned out that its value (\ref{c10}) can be represented by symmetric invariants of the permutation group $S_4$ acting on the Euclidean space ${\mathbb E}^4$ with following equating two variables (\ref{e27}). That is why we extended our algebraic approach to a four-phase symmetric composite. 

A further extension of the $n$-color tiling, preserving the isotropy of the second-rank tensor, to $n\geq 4$, $n\neq 6$, arises in the quasicrystalline color tilings. These tilings arise either in irrational projections of the perfect high-dimensional lattice onto the plane \cite{lif97}, e.g., the five-color Penrose tiling, or in hierarchical tilings of the plane \cite{rad94}, e.g., the five-color pinwheel tiling.

There is another topological property, {\em percolation}, which is related to the problem of the minimal number of insulating phases that make the entire compound an insulator. Specifically, we refer to the existence of {\em  monochromatic} ({\em M}) {\em  loops} in color checkerboards (see the first three checkerboards in Figure \ref{fig3}). We did not find similar structures with these {\em M}- loops among regular color checkerboards with $n\geq 5$. This is not accidental, but reflects a basic topological property of color tiling on the plane and is related to the celebrated four-color problem.

We illustrate how the universal properties of the isotropic effective conductivity $\sigma_e(\sigma_1,\ldots,\sigma_n)$ in a 2D $n$-phase composite with a rotation axis $C_r$, where $r\ge 3$, diminish with increasing $n$:
\bea
\left\{\begin{array}{c}
n=2\\{\sf h}_2={\sf C}_6\\{\sf P}_2=S_2\\M-loop\\
\sigma_e=\sqrt{\sigma_1\;\sigma_2}\end{array}\right\}
\longrightarrow\left\{\begin{array}{c}
n=3\\{\sf h}_3={\sf C}_3\\{\sf P}_3=S_3\\M-loop\\-
\end{array}\right\}
\longrightarrow\left\{\begin{array}{c}
n=4\\{\sf h}_4={\sf C}_4\\{\sf P}_4=D_2\\M-loop\\-
\end{array}\right\}
\longrightarrow\left\{ \begin{array}{c}
n=6\\{\sf h}_6={\sf C}_3\\{\sf P}_6=D_3\\-\\-\end{array}
\right\}\nonumber
\eea
where a sign "--" in the forth and fifth lines marks an absence of {\em M}-loops and universal solutions, respectively.
%%%%%%%%%%%%%%%%%%%%%%%%%%%%%%%%%%
%%%%%%%%%%%%%%%%%%%%%%%%%%%%%%%%%%
\section{Effective conductivity $\sigma_e(\sigma_1,\ldots,\sigma_n)$ and its algebraic properties}\label{part3}
%%%%%%%%%%%%%%%%%%%%%%%%%%%%%%%%%%%
%%%%%%%%%%%%%%%%%%%%%%%%%%%%%%%%%%
In this section, we list the main properties of isotropic effective conductivity $\sigma_e(\sigma_1,\ldots,\sigma_n)$ of the 2D $n$-phase composites that are derived from physics. A direct way to solve the EC problem starts with the local field equations for electrical fields ${\bf E}({\bf r})$ and current density ${\bf J}({\bf r})$
\bea
\nabla\times\;{\bf E}({\bf r})=0,\qquad\nabla\cdot\;{\bf J}({\bf r})=0,\qquad{\bf J}({\bf r})=\sigma({\bf r})\;{\bf E}({\bf r}),\label{c1}
\eea
along with appropriate boundary conditions for the electrical potential. The local isotropic conductivity $\sigma ({\bf r})$ is a discontinuous function, defined as $\sigma({\bf r})=\sigma_i$ if ${\bf r}\in\Delta_i\;,\;i=1,\ldots,n$, where $\Delta_i$ is a homogeneous part of the compound with constant conductivity $\sigma
_i$. The isotropic effective conductivity $\sigma_e=\sigma_e(
\sigma_1,\ldots,\sigma_n)$ can be defined by Ohm's law for the system's averaged current density ${\bf J}_e$ and the field ${\bf E}_e$,
\bea
{\bf J}_e=\sigma_{e}\;{\bf E}_e,\qquad{\bf J}_e=\frac{1}{S}
\int_S{\bf J}({\bf r})\;d S,\qquad{\bf E}_e=\frac{1}{S}
\int_S{\bf E}({\bf r})\;d S.\label{c2}
\eea
List the main properties of the function $\sigma_{e}(\sigma_1,\ldots,\sigma_n)$, which are derived from physical considerations.

%%%%%%%%%%%%%%%%%%%%%%%%%%%%%%%%%%%%%%%%%%%
$\bullet$ 
{\sf Homogeneity of 1st order}
\bea
\sigma_e(k\sigma_1,\ldots,k\sigma_n)=k\;\sigma_e(\sigma_1,\ldots,\sigma_n).\label{c3}
\eea
It follows from the linearity of equations (\ref{c1}) and the definitions of the average current and field (\ref{c2}).

$\bullet$
{\sf Permutation invariance}
\bea
\sigma_e({\widehat {\cal P}_l}\;\{\sigma_1,\ldots,\sigma_n\})=\;\sigma_e(\sigma_1,\ldots,\sigma_n)\;,\label{c4}
\eea
where ${\widehat {\cal P}_l}$ permutates the indices $\{1,\ldots,
n\}$. The operators ${\widehat {\cal P}_l}$ form a symmetric group $S_n$. The existence of this permutation invariance implies a distribution of $n$ phases with equal volume fraction $p=1/n$.

$\bullet$
{\sf Self-duality \cite{fel00}}
\bea
\sigma_e(\sigma_1,\ldots,\sigma_n) \times\sigma_e(\sigma_1^{-1},
\ldots,\sigma_n^{-1})=1.\label{c5}
\eea

$\bullet$
{\sf Compatibility}
\bea
\sigma_e(\sigma,\ldots,\sigma)=\sigma.\label{c6}
\eea
This follows from the Maxwell equations (\ref{c1}) and the definition (\ref{c2}) of $\sigma_e(\sigma_1,\ldots,\sigma_n)$.

$\bullet$
{\sf Positivity and Monotony}
\bea
\sigma_e(\sigma_1,\ldots,\sigma_n)>0,\qquad\frac{\partial
\sigma_e(\sigma_1,\ldots,\sigma_n)}{\partial \sigma_i}>0,\quad i=1,2,3.\label{c7}
\eea
The first inequality follows from physics while the second one follows from the first-order homogeneity of the function $\sigma_e(\sigma_1,\ldots,\sigma_n)$. Indeed, according to the equivalent definition of EC
$$
\sigma_e(\sigma_1,\ldots,\sigma_n)=\frac{1}{SE_e^2}\sum_{i=1}^n
\sigma_i\int_{s_i}{\bf E}^2({\bf r})ds,\qquad S=\sum_{i=1}^n s_i,
$$
where $s_i$ denotes a surface measure of $\Delta_i$. By Euler's theorem for the homogeneous function of the first order, this leads to the following
$$
\sigma_e(\sigma_1,\ldots,\sigma_n)=\sum_{i=1}^n\sigma_i\frac{
\partial\sigma_e}{\partial\sigma_i}\qquad\rightarrow\qquad
\frac{\partial\sigma_e(\sigma_1,\ldots,\sigma_n)}{\partial 
\sigma_i}=\frac{1}{SE_e^2}\int_{s_i}{\bf E}^2({\bf r})ds>0.
$$
Inequalities (\ref{c7}) are strict for all phases with finite volume fractions $p_i=s_i/S$.

$\bullet$
{\sf Dykhne's ansatz (exact solution) \cite{dyk70}}

\noindent
If there exists such a $\sigma_*$ that for every $\sigma_i$ one can point out such a $\sigma_j$, which satisfies
\bea
\sigma_i\cdot\sigma_j=\sigma_*^2,\qquad\mbox{then}\qquad
\sigma_e(\sigma_1,\ldots,\sigma_n)=\sigma_*=
\prod_{i=1}^{n}\sqrt[n]{\sigma_i}\;,\label{c8}
\eea
%%%%%%%%%%%%%%%%%%%%%%%%%%%%%%%%%%%%%%%%%%%
%%%%%%%%%%%%%%%%%%%%%%%%%%%%%%%%%%%%%%%%%%%
\subsection{Approximation, exact solution and asymptotics}\label{part31}
%%%%%%%%%%%%%%%%%%%%%%%%%%%%%%%%%%%%%%%%%%%
%%%%%%%%%%%%%%%%%%%%%%%%%%%%%%%%%%%%%%%%%%%
$\bullet$
{\sf Bruggeman's equation ($n$-phase composite) \cite{bru35}}
\bea
\sum_{j=1}^n\frac{\sigma_{Br}(\sigma_1,\ldots,\sigma_n)-\sigma_j}{\sigma_{Br}(\sigma_1,\ldots,\sigma_n)+\sigma_j}=0\;.\label{c9}
\eea
The Bruggeman approximation $\sigma_{Br}(\sigma_1,\ldots,\sigma_n)$ provides a remarkably accurate approximation for EC of the 2D random multi-phase lattice wire system. It has been shown \cite{luc91,felk02} that in the case of the square lattice the first four orders of the expansion of Bruggeman's solution in terms of the moments of the disorder parameter coincide with the corresponding terms of the expansion of the exact solution. However, in the 5th order the Bruggeman approximation deviates from the exact one.
%%%%%%%%%%%%%%%%%%%%%%%%%%%%%%%%%%%%%%%%%%%

$\bullet$
{\sf Exact solution $\sigma_{MS}^{ij}$ for the Mortola-Steff\'e checkerboard in four phases (Fig. \ref{fig2}) \cite{mor85,mil01,cra01}}

\noindent
The anisotropic conductivity tensor $\sigma_{MS}^{ij}(\sigma_1,\sigma_2,\sigma_3,\sigma_4)$ has the principal values,
\bea
\sigma_{MS}^{xx}(\sigma_1,\sigma_2,\sigma_3,\sigma_4)=\sqrt{\frac{
(\sigma_1+\sigma_2)(\sigma_3+\sigma_4)}{(\sigma_1+\sigma_4)
(\sigma_2+\sigma_3)}}\;\sqrt{\frac{\sigma_1\sigma_2\sigma_3
+\sigma_1\sigma_2\sigma_4+\sigma_1\sigma_3\sigma_4+\sigma_2
\sigma_3\sigma_4}{\sigma_1+\sigma_2+\sigma_3+\sigma_4}},\nonumber\\
\sigma_{MS}^{yy}(\sigma_1,\sigma_2,\sigma_3,\sigma_4)=
\sqrt{\frac{(\sigma_1+\sigma_4)(\sigma_2+\sigma_3)}
{(\sigma_1+\sigma_2)(\sigma_3+\sigma_4)}}\;
\sqrt{\frac{\sigma_1\sigma_2\sigma_3+\sigma_1\sigma_2\sigma_4+
\sigma_1\sigma_3\sigma_4+\sigma_2\sigma_3\sigma_4}
{\sigma_1+\sigma_2+\sigma_3+\sigma_4}}.\nonumber
\eea
By equating the partial conductivities $\sigma_2=\sigma_4$, the tensor $\sigma_{MS}^{ij}$ becomes isotropic,
\bea
\sigma_{MS}(\sigma_1,\sigma_2,\sigma_3,\sigma_2)=\sqrt{\sigma_2\;\frac{2\sigma_1\sigma_3+\sigma_2(\sigma_1+\sigma_3)}
{\sigma_1+2\sigma_2+\sigma_3}}\;.\label{c10}
\eea
%%%%%%%%%%%%%%%%%%%%%%%%%%%%%%%%%%%%%%%%%%%

$\bullet$
{\sf Asymptotics of $\sigma_{Rh}(\sigma_1,\sigma_2,\sigma_3)$ for a three-phase rhombic compound {\sf Rh} (Fig. \ref{fig1}d) \cite{kha00}}

\noindent
Applying an approach developed by Keller \cite{kel87} for constituents of large contrast in their conductivities, the leading terms of the asymptotic expansion for $\sigma_{Rh}(\sigma_1,\sigma_
2,\sigma_3)$ were derived \cite{kha00} in two cases,
\bea
\sigma_1\!<<\!\sigma_2,\sigma_3,\quad\sigma_{Rh}\!\simeq\!\sqrt{\sigma_1}\;\left(\sqrt{\sigma_2}+
\sqrt{\sigma_3}\right),\qquad\sigma_1\!>>\!\sigma_2=\sigma_3,\quad\sigma_{Rh}\simeq\frac{1}{2}\;
\sqrt{\sigma_1\;\sigma_2}.\label{c11}
\eea
%%%%%%%%%%%%%%%%%%%%%%%%%%%%%%%%%%%%%%%%%%%
%%%%%%%%%%%%%%%%%%%%%%%%%%%%%%%%%%%%%%%%%%%
\subsection{Lower and upper bounds for $\sigma_e(\sigma_1,\ldots,
\sigma_n)$}\label{part32}
%%%%%%%%%%%%%%%%%%%%%%%%%%%%%%%%%%%%%%%%%%%
%%%%%%%%%%%%%%%%%%%%%%%%%%%%%%%%%%%%%%%%%%%
$\bullet$
{\sf Wiener's bounds ($n$-phase composite) \cite{wie12}}
\bea
\sigma_{Wi}^{-}(\sigma_1,\ldots,\sigma_n)=\left(\frac{1}{n}\sum_{i=
1}^n\frac{1}{\sigma_i}\right)^{-1}\leq\sigma_e(\sigma_1,\ldots,
\sigma_n)\leq\frac{1}{n} \sum_{i=1}^n\sigma_i=\sigma_{Wi}^{+}(\sigma_1,\ldots,\sigma_n).\label{c12}
\eea

\noindent
Bounds $\sigma_{Wi}^+(\sigma_1,\ldots,\sigma_n)$ and $\sigma_{Wi}^-
(\sigma_1,\ldots,\sigma_n)$ are permutation-invariant and mutually (not self-) dual
\bea
\sigma_{Wi}^{\pm}(\sigma_1,\ldots,\sigma_n)\times\sigma_{Wi}^{\mp}\left(\sigma_1^{-1},\ldots,\sigma_n^{-1}\right)=1.\label{c13}
\eea
Wiener's bounds do not satisfy ansatz (\ref{c8}). In fact, they are valid for the eigenvalues of the tensor $\sigma_e^{ij}$ of any anisotropic structure. 
%%%%%%%%%%%%%%%%%%%%%%%%%%%%%%%%%%%%%%%%%%%

$\bullet$
{\sf Hashin-Shtrikman's bounds ($n$-phase composite) \cite{has62}}
\bea
\sigma_{HS}^-(\sigma_1,\ldots,\sigma_n)\leq\sigma_{e}(\sigma_1,
\ldots,\sigma_n)\leq\sigma_{HS}^+(\sigma_1,\ldots,\sigma_n),\label{c14}
\eea
where $\sigma_{HS}^+$ and $\sigma_{HS}^-$ are determined for the ordered sequence of  partial conductivities $\sigma_1\leq\ldots
\leq\sigma_n$,
\bea
\sigma_{HS}^-=\sigma_1\frac{1+A_1}{1-A_1},\quad A_1=\frac{1}{n}\sum_{i=2}^n\frac{\sigma_i-\sigma_1}{\sigma_i+\sigma_1},\quad
\sigma_{HS}^+=\sigma_n\frac{1+A_n}{1-A_n},\quad A_n=\frac{1}{n}
\sum_{i=1}^{n-1}\frac{\sigma_i-\sigma_n}{\sigma_i+\sigma_n}.\quad
\label{c15}
\eea
The bounds $\sigma_{HS}^+(\sigma_1,\ldots,\sigma_n)$ and $\sigma_
{HS}^-(\sigma_1,\ldots,\sigma_n)$ are self-dual
$$
\sigma_{HS}^{\pm}(\sigma_1,\ldots,\sigma_n)\times\sigma_{HS}^{\pm}
\left(\sigma_1^{-1},\ldots,\sigma_n^{-1}\right)=1,
$$
but do not satisfy the permutation invariance (\ref{c4}) and therefore do not satisfy the ansatz (\ref{c8}).
%%%%%%%%%%%%%%%%%%%%%%%%%%%%%%%%%%%%%%%%%%%

$\bullet$
{\sf Nesi's bounds (three- phase composite) \cite{nes91,nes95}}

\noindent
Applying the null-Lagrangian method, new bounds $\sigma_{Ne}^+$ and $\sigma_N^-$ were derived for $\sigma_e(\sigma_1,\sigma_2,\sigma_3)$ of the isotropic 2D three-phase composite made of cyclically symmetric isotropic constituents,
\bea
\sigma_{Ne}^-(\sigma_1,\sigma_2,\sigma_3)\leq\sigma_e(\sigma_1,\sigma_2,\sigma_3)\leq\sigma_{Ne}^+(\sigma_1,\sigma_2,\sigma_3).\label{c16}
\eea

The bounds $\sigma_{Ne}^{\pm}(\sigma_1,\sigma_2,\sigma_3)$ are the unique positive roots of the cubic equations
\bea
\det\left(\begin{array}{ccc}
2\sigma_1 & -\sigma_{Ne}^- & -\sigma_{Ne}^-\\
-\sigma_{Ne}^- & 2\sigma_2 & -\sigma_{Ne}^-\\
-\sigma_{Ne}^- & -\sigma_{Ne}^-& 2\sigma_3\end{array}\right)=0,\qquad
\det\left(\begin{array}{ccc}
2/\sigma_1 & -1/\sigma_{Ne}^+ & -1/\sigma_{Ne}^+\\
-1/\sigma_{Ne}^+ & 2/\sigma_2 & -1/\sigma_{Ne}^+\\
-1/\sigma_{Ne}^+ & -1/\sigma_{Ne}^+ & 2/\sigma_3\end{array}\right)=0\;,\label{c17}
\eea
or, more simply
\bea
\left(\sigma_{Ne}^-\right)^3+I_1\;\left(\sigma_{Ne}^-\right)^2-4\;
I_3=0,\qquad\left(\sigma_{Ne}^+\right)^3-\frac{1}{4}\;I_2\;
\sigma_{Ne}^+-\frac{1}{4}\;I_3=0,\qquad\mbox{where}\label{c18}\\
I_1=\sigma_1+\sigma_2+\sigma_3,\quad I_2=\sigma_1\sigma_2+\sigma_2
\sigma_3+\sigma_3\sigma_1,\quad I_3=\sigma_1\sigma_2\sigma_3.
\nonumber
\eea
%%%%%%%%%%%%%%%%%%%%%%%%%%%%%%%%%%%%%%%%%%%

$\bullet$
{\sf Conjectured bounds (four-phase composite)}

\noindent
Regarding the EC problem in the 2D n-phase composites with 
arbitrary $n\ge 3$, neither the paper \cite{nes91} nor \cite{nes95} provides the equations in the closed forms or explicit expressions for the bounds at $n\ge 4$, but only for $n=3$ \cite{nes25}. That is why, based on the matrix representation (\ref{c17}) when $n\!=\!3$ and exact solution $\sigma_{Kl}^{\pm}$ when $n\!=\!2$ with the matrix representations (a subscript '{\em Kl} ' stands for J. Keller)
$$
\det\left(\begin{array}{cc}\sigma_1&-\sigma_{Kl}^-\\-\sigma_{Kl}^-&\sigma_2\end{array}\right)=0,\qquad\det\left(\begin{array}{cc}1/
\sigma_1& -1/\sigma_{Kl}^+\\-1/\sigma_{Kl}^+& 1/\sigma_2\end{array}\right)=0,\qquad\sigma_{Kl}^{\pm}=\sqrt{\sigma_1\sigma_2},
$$ 
we suggest here for the 2D four-phase composites the following determinant equations 
\bea
\det\left(\begin{array}{cccc}
3\sigma_1 & -\sigma_{Cj}^- & -\sigma_{Cj}^-& -\sigma_{Cj}^-\\
-\sigma_{Cj}^- & 3\sigma_2 & -\sigma_{Cj}^-& -\sigma_{Cj}^-\\
-\sigma_{Cj}^- & -\sigma_{C_4}^-& 3\sigma_3& -\sigma_{Cj}^-\\
-\sigma_{Cj}^- & -\sigma_{Cj}^-& -\sigma_{Cj}^-& 3\sigma_4
\end{array}\right)\!=\!0,\qquad
\det\left(\begin{array}{cccc}
3/\sigma_1& -1/\sigma_{Cj}^+&-1/\sigma_{Cj}^+&-1/\sigma_{Cj}^+\\
-1/\sigma_{Cj}^+ &3/\sigma_2&-1/\sigma_{Cj}^+&-1/\sigma_{Cj}^+\\
-1/\sigma_{Cj}^+ &-1/\sigma_{Cj}^+&3/\sigma_3&-1/\sigma_{Cj}^+\\
-1/\sigma_{Cj}^+ &-1/\sigma_{Cj}^+&-1/\sigma_{Cj}^+ & 3/\sigma_4
\end{array}\right)\!=\!0\;,\nonumber
\eea
which are converted into two quartic equations for the bounds  
$\sigma_{Cj}^{\pm}(\sigma_1,\sigma_2,\sigma_3,\sigma_4)$
\bea
\left(\sigma_{Cj}^-\right)^4+2J_1\left(\sigma_{Cj}^-\right)^3+
3J_2\left(\sigma_{Cj}^-\right)^2-27J_4=0,\label{c19}
\eea
$$
\left(\sigma_{Cj}^+\right)^4-\frac1{9}J_2\left(\sigma_{Cj}^+\right)^2-\frac{2}{27}J_3\;\sigma_{Cj}^+-\frac1{27}J_4=0,\qquad
\mbox{where}
$$
$$
J_1=\sum_{j=1}^4\sigma_j,\quad J_2=\sum_{k\ge j=1}^4
\sigma_j\sigma_k,\quad J_3=\sum_{r\ge k\ge j=1}^4\sigma_j\sigma_k
\sigma_r,\quad J_4=\prod_{j=1}^4\sigma_j.
$$

\noindent
Conjectured bounds are permutation invariant and mutually dual
\bea
\sigma_{Cj}^-(\sigma_1,\sigma_2,\sigma_3,\sigma_4)\cdot
\sigma_{Cj}^+(\sigma_1^{-1},\sigma_2^{-1},\sigma_3^{-1},
\sigma_4^{-1})=1.\nonumber
\eea
In Section \ref{part61} we justify this conjecture by comparison with the other known bounds.
%%%%%%%%%%%%%%%%%%%%%%%%%%%%%%%%%%%%%%%%%%%

$\bullet$
{\sf Cherkaev bounds (three-phase composite with one superconducting phase) \cite{che09,che12,che24}}

\noindent
Sophisticated expressions for the extremal values of the isotropic effective conductivity were explicitly determined in \cite{che09} for the ordered phase conductivities, $\sigma_1\le\sigma_2\le
\sigma_3$,
\bea
\sigma_{Ch}^-(\sigma_1,\sigma_2,\sigma_3)\le 
\sigma_e(\sigma_1,\sigma_2,\sigma_3)\le
\sigma_{Ch}^+(\sigma_1,\sigma_2,\sigma_3).\label{c20}
\eea
Due to the great technical difficulties in analyzing these expressions, we consider here a simple case \cite{che12,che24} when one of the phases is {\em almost superconducting}, 
\bea
\sigma_1\le\sigma_2\ll\sigma_3.\label{c21}
\eea
We present the formulas (see (9 -- 13) in \cite{che12} and (30,31) in \cite{che24}) for the lower bound $\sigma_{Ch}^-\left(\sigma_1,\sigma_2,\sigma_3\right)$, which are adapted to a three-phase isotropic compound with equal volume fraction by assumption (\ref{c21}),
%%%%%%%%%%%%%%%%%%%%%%%%%%%%%%%%%%%%%%%%%%%
\bea
\sigma_{Ch}^-\left(\sigma_1,\sigma_2,\sigma_3\right)
=\left\{\begin{array}{ccl}
U_1, & if & p_{11}\le\frac{1}{3}\\
U_2, & if & p_{12}\le\frac{1}{3}\le p_{11}\\
U_3, & if & \frac{1}{3}\le p_{12}\end{array}\right.,\quad
\left.\begin{array}{ccl}
U_1\!\!\!&=&\!\!\!-\sigma_1+3\left(\frac1{2\sigma_1}+\frac1{
\sigma_1+\sigma_2}\right)^{-1},\\
U_2\!\!\!&=&\!\!\!\sigma_2+6\sigma_1\left(1-\frac1{\sqrt{3}}\right)^2,\\
U_3\!\!\!&=&\!\!\!-\sigma_2+6\left(\frac1{\sigma_1}+\frac1{
\sigma_2}\right)^{-1},\end{array}\right.\label{c22}
\eea
where
\bea
p_{11}=\frac{\sqrt{3}-1}{3}\frac{2\sigma_1}{\sigma_1+\sigma_2},\qquad 
p_{12}=\frac{\sqrt{3}-1}{3}\frac{\sigma_1}{\sigma_2},\qquad p_{11}\ge p_{12}.\label{c23}
\eea
A simple analysis shows that the last two options in (\ref{c22}) are forbidden. Indeed, 
\begin{itemize}
\item
If $1/3\le p_{12}$, then $\sigma_2/\sigma_1\le\sqrt{3}-1\simeq 0.73
$, which contradicts the ascending order (\ref{c21}) in $\sigma_j$.
\item
If $p_{12}\le 1/3\le p_{11}$, then $\sqrt{3}-1\le\sigma_2/\sigma_1
\le  2\sqrt{3}-3$, which again leads to a contradiction.
\end{itemize}
Consider the first option in (\ref{c22}) and obtain
\bea
\frac{\sigma_2}{\sigma_1}\ge 2\sqrt{3}-3,\qquad\sigma_{Ch}^-\left(
\sigma_1,\sigma_2,\sigma_3\right)=\sigma_1\frac{3\sigma_1+5\sigma_
2}{3\sigma_1+\sigma_2}.\label{c24}
\eea

The upper bound $\sigma_{Ch}^+\left(\sigma_1,\sigma_2,\sigma_3
\right)$ is dual to its lower bound (see formula (6.3) in \cite{che09}),
\bea
\sigma_{Ch}^+\left(\sigma_1,\sigma_2,\sigma_3\right)\cdot\sigma_
{Ch}^-\left(\sigma_3^{-1},\sigma_2^{-1},\sigma_1^{-1}\right)=1.
\label{c25}
\eea
To calculate $\sigma_{Ch}^-\left(\sigma_3^{-1},\sigma_2^{-1},\sigma_1^{-1}\right)$ we again make use of the formulas in \cite{che12} and \cite{che24}, bearing in mind $\sigma_3^{-1}\ll
\sigma_2^{-1}\le\sigma_1^{-1}$,
\bea
\sigma_{Ch}^-\left(\sigma_3^{-1},\sigma_2^{-1},\sigma_1^{-1}\right)
=\left\{\begin{array}{ccl}
U_1^*, & if & p_{11}^*\le\frac{1}{3}\\
U_2^*, & if & p_{12}^*\le\frac{1}{3}\le p_{11}\\
U_3^*, & if & \frac{1}{3}\le p_{12}^*\end{array}\right.,\;
\left.\begin{array}{ccl}
U_1^*\!\!&=&\!\!-\sigma_3^{-1}+3\left(\frac1{2\sigma_3^{-1}}+
\frac1{\sigma_3^{-1}+\sigma_2^{-1}}\right)^{-1},\\
U_2^*\!\!&=&\!\!\sigma_2^{-1}+6\sigma_3^{-1}\left(1-\frac1
{\sqrt{3}}\right)^2,\\
U_3^*\!\!&=&\!\!-\sigma_2^{-1}+6\left(\frac1{\sigma_3^{-1}}+\frac1{\sigma_2^{-1}}\right)^{-1},\end{array}\right.\nonumber
\eea
where
\bea
p_{11}^*=\frac{\sqrt{3}-1}{3}\frac{2\sigma_3^{-1}}{\sigma_3^{-1}+
\sigma_2^{-1}},\qquad p_{12}^*=\frac{\sqrt{3}-1}{3}\frac{\sigma_3^
{-1}}{\sigma_2^{-1}},\qquad p_{11}^*\ge p_{12}^*.\nonumber
\eea
A simple analysis, much like the previous one, yields
\bea
\frac{\sigma_2^{-1}}{\sigma_3^{-1}}\ge 2\sqrt{3}-3,\qquad
\sigma_{Ch}^-\left(\sigma_3^{-1},\sigma_2^{-1},\sigma_1^{-1}\right)
=\frac{5+3\nu}{1+3\nu}\;\frac1{\sigma_3},\qquad
\nu=\frac{\sigma_2}{\sigma_3}\ll 1.\nonumber
\eea
Combining the last formula with (\ref{c25}) and neglecting the small terms in $\nu$, we obtain
\bea
\sigma_{Ch}^+\left(\sigma_1,\sigma_2,\sigma_3\right)
\simeq \frac1{5}\;\sigma_3.\label{c26}
\eea
%%%%%%%%%%%%%%%%%%%%%%%%%%%%%%%%
%%%%%%%%%%%%%%%%%%%%%%%%%%%%%%%%%
\section{Self-dual symmetric polynomials}\label{part4}
%%%%%%%%%%%%%%%%%%%%%%%%%%%%%%%%
%%%%%%%%%%%%%%%%%%%%%%%%%%%%%%%
In this section, we recall the main definitions and basic properties \cite{fel22} of self-dual multivariate  symmetric polynomials in 
$\lambda$ with coefficients built upon the basic polynomial invariants $I_{n,r}({\bf x}^n)=I_{n,r}$ of the symmetric group $S_n$, acting on the Euclidean space ${\mathbb E}^n$
\bea
I_{n,r}&=&\sum_{i_1<i_2<\ldots<i_r}^{n} x_{i_1}x_{i_2}\ldots x_{i_r},\qquad{\bf x^n}=\{x_1,\ldots,x_n\}
\in{\mathbb E}^n,\quad\mbox{i.e.}\label{d1}\\
I_{n,0}=1,\quad I_{n,1}\!\!\!&=&\!\!\!\sum_{i}^n x_i,\quad I_{n,2}=
\sum_{i<j}^nx_ix_j,\quad\ldots,\quad I_{n,n-1}=I_{n,n}\;\sum_{i}^n
\frac{1}{x_i},\quad I_{n,n}=\prod_{i}^n x_i.\quad\nonumber
\eea
These polynomials exhibit a generalization of {\em univariate self-dual} polynomials ${\sf R}_m(\lambda)$ and ${\sf S}_m(\lambda)$
\bea 
{\sf R}_m(\lambda)=\sum_{k=0}^{\left\lfloor m/2\right\rfloor}{\cal R}_k^m\left(\lambda^{m-k}+\lambda^k\right),\qquad
{\sf S}_m(\lambda)=\sum_{k=0}^{\left\lfloor m/2\right\rfloor}{\cal S}_k^m\left(\lambda^{m-k}-\lambda^k\right),\label{d2}
\eea
satisfying the duality relations
\bea
\lambda^m{\sf R}_m(\lambda^{-1})-{\sf R}_m(\lambda)=0,\qquad
\lambda^m{\sf S}_m(\lambda^{-1})+{\sf S}_m(\lambda)=0.\label{d3}
\eea
Define {\em multivariate} self-dual polynomials in $\lambda$ of degree $mn$, $m\in{\mathbb Z}_{\ge}$ and $n\ge 2$,
\bea
\hspace{-1cm}
&&{\sf R}\left(^{\lambda,\;{\bf x^n}}_{m,S_n}\right)=\sum_{s=0}^
{\left\lfloor s_*\right\rfloor}\left[R_s^+\left(^{\bf x^n}_{m,S_n}
\right)\lambda^{mn-s}+R_s^-\left(^{\bf x^n}_{m,S_n}\right)
\lambda^s\right],\qquad R_0^-\left(^{\bf x^n}_{m,S_n}\right)=
R_0^+\left(^{\bf x^n}_{m,S_n}\right)I_{n,n}^m,\label{d4}\\
\hspace{-1cm}
&&{\sf S}\left(^{\lambda,\;{\bf x^n}}_{m,S_n}\right)=
\sum_{s=0}^{\left\lfloor s_*\right\rfloor}\left[S_s^+
\left(^{\bf x^n}_{m,S_n}\right)\lambda^{mn-s}-S_s^-
\left(^{\bf x^n}_{m,S_n}\right)\lambda^s\right],\qquad 
S_0^-\left(^{\bf x^n}_{m,S_n}\right)=S_0^+\left(^{\bf x^n}_
{m,S_n}\right)I_{n,n}^m.\label{d5}
\eea
Here, $\left\lfloor x\right\rfloor$ denotes the floor function, $s_*=\frac{mn}{2}$ and the symbols $\left(^{\lambda,\;{\bf x^n}}_
{m,S_n}\right)$ and $\left(^{\bf x^n}_{m,S_n}\right)$ denote a dependence of the polynomials on their entries. The summands $R_s^{\pm}\left(^{\bf x^n}_{m,S_n}\right)$ and $S_s^{\pm}\left(^
{\bf x^n}_{m,S_n}\right)$ in (\ref{d4},\ref{d5}) read
\bea
\hspace{-.6cm}
&&R_s^+\left(^{\bf x^n}_{m,S_n}\right)\!=\!\sum_{l=1}^
{P_n^m(s)}{\cal R}^{s,\;l}_{m,n}\prod_{r=0}^nI_{n,r}^
{\alpha_{n-r}\left(^{mn-s,\;\;l}_{m,\hspace{.5cm}S_n}\right)},
\qquad R_s^-\left(^{\bf x^n}_{m,S_n}\right)\!=\!
\sum_{l=1}^{P_n^m(s)}{\cal R}^{s,\;l}_{m,n}\prod_{r=0}^nI_{n,n-r}^
{\alpha_r\left(^{s,\;\;l}_{m, S_n}\right)},\qquad\label{d6}\\
\hspace{-.6cm}
&&S_s^+\left(^{\bf x^n}_{m,S_n}\right)\!=\!\sum_{l=1}^
{P_n^m(s)}{\cal S}^{s,\;l}_{m,n}\prod_{r=0}^nI_{n,r}^
{\alpha_{n-r}\left(^{mn-s,\;\;l}_{m,\hspace{.5cm}S_n}\right)},
\qquad S_s^-\left(^{\bf x^n}_{m,S_n}\right)\!=\!
\sum_{l=1}^{P_n^m(s)}{\cal S}^{s,\;l}_{m,n}\prod_{r=0}^n
I_{n,n-r}^{\alpha_r\left(^{s,\;\;l}_{m, S_n}\right)},\label{d7}\\
\hspace{-.6cm}
&&\alpha_r\left(^{s,\;\;l}_{m, S_n}\right)=\alpha_{n-r}
\left(^{mn-s,\;\;l}_{m,\hspace{.5cm}S_n}\right),\quad
\alpha_r\left(^{s,\;\;l}_{m,S_n}\right)\in{\mathbb Z}_{\ge},
\qquad {\cal R}^{s,\;l}_{m,n},\;{\cal S}^{s,\;l}_{m,n}\in
{\mathbb R}_{\ge},\quad P_n^m(0)=1.\label{d8}
\eea
In formulas (\ref{d6}, \ref{d7}), the function $P_n^m(s)$ denotes a partition number with constraints. Specifically, $P_n^m(s)$ gives the number of non-negative integer solutions $\left\{\alpha^l\right\}$ of two Diophantine equations,
\bea
\sum_{r=0}^nr\;\alpha_r\left(^{s,\;\;l}_{m, S_n}\right)=s,
\qquad\sum_{r=0}^n\alpha_r\left(^{s,\;\;l}_{m,S_n}\right)=m,
\qquad\alpha_r\left(^{0,\;\;l}_{m,S_n}\right)=m\delta_{ r,0},\label{d9}
\eea
which are solvable if $0\le s\le mn$ (a detailed description see in \cite{fel22}, Appendix B). 

Polynomials ${\sf R}\left(^{\lambda,\;{\bf x^n}}_{m,S_n}\right)$ and ${\sf S}\left(^{\lambda,\;{\bf x^n}}_{m,S_n}\right)$ coincide with univariate polynomials ${\sf R}_m(\lambda)$ and ${\sf S}_m(
\lambda)$ if $n=1$ and they satisfy the duality relations
\bea
\lambda^{mn}I_{n,n}^m{\widehat{\bf Q}_n}\;{\sf R}\left(^{
\lambda,\;{\bf x^n}}_{m,S_n}\right)-{\sf R}\left(^{\lambda,\;
{\bf x^n}}_{m,S_n}\right)=0,\qquad
\lambda^{mn}I_{n,n}^m{\widehat{\bf Q}_n}\;{\sf S}\left(^{
\lambda,\;{\bf x^n}}_{m,S_n}\right)+{\sf S}\left(^{\lambda,\;
{\bf x^n}}_{m,S_n}\right)=0,\label{d10}
\eea
generalizing (\ref{d3}). The operator ${\widehat{\bf Q}_n}$ in (\ref{d10}) acts on the polynomial ${\sf H}\left(^{\lambda,\;
{\bf x^n}}_{m,S_n}\right)$ by inverting $\lambda$ and $x_i$,
\bea
{\widehat{\bf Q}_n}\;{\sf H}\left(^{\lambda,\;{\bf x^n}}_{m,S_n}
\right)={\sf H}\left(^{\lambda^{-1},\;{\bf x^{-n}}}_{m,\;\;\;S_n}\right),\qquad 
{\widehat{\bf Q}_n}I_{n,r}\left({\bf x^n}\right)=I_{n,r}\left({\bf x^{-n}}\right),\quad{\bf x^{-n}}=\left\{\frac1{x_1},\ldots,
\frac1{x_n}\right\}.\label{d11}
\eea
As an instructive example, we present self-dual polynomials with small $n\le 4$ and $m=1,2$.
{\small
\bea
&&\hspace{-.8cm}
{\sf S}\left(^{\lambda,\;{\bf x^2}}_{1,\;S_2}\right)=
\;\;\lambda^2-I_{2,2},\qquad
{\sf R}\left(^{\lambda,\;{\bf x^2}}_{1,\;S_2}\right)=
L_1\lambda^2+L_2I_{2,1}\lambda+L_1I_{2,2}^2,\label{d12}\\
&&\hspace{-.8cm}
{\sf S}\left(^{\lambda,\;{\bf x^2}}_{2,\;S_2}\right)=
K_1\lambda^4+K_2I_{2,1}\lambda^3-K_2I_{2,1}I_{2,2}\lambda
-K_1I_{2,2}^2,\nonumber\\
&&\hspace{-.8cm}
{\sf R}\left(^{\lambda,\;{\bf x^2}}_{2,\;S_2}\right)=
L_1\lambda^4+L_2I_{2,1}\lambda^3+\left(L_3I_{2,1}^2+L_4I_{2,2}\right)
\lambda^2+L_2I_{2,1}I_{2,2}\lambda+L_1I_{2,2}^2,\nonumber\\
&&\hspace{-.8cm}
{\sf S}\left(^{\lambda,\;{\bf x^3}}_{1,\;S_3}\right)=
K_1\lambda^3\!+\!K_2I_{3,1}\lambda^2\!-\!K_2I_{3,2}\lambda\!-\!K_1I_{3,3},\qquad
{\sf R}\left(^{\lambda,\;{\bf x^3}}_{1,\;S_3}\right)\!=\!L_1\lambda^3\!+\!L_2I_{3,1}
\lambda^2\!+\!L_2I_{3,2}\lambda\!+\!L_1I_{3,3},\nonumber\\
&&\hspace{-.8cm}
{\sf S}\left(^{\lambda,\;{\bf x^4}}_{1,\;S_4}\right)=
K_1\lambda^4\!+\!K_2I_{4,1}\lambda^3\!-\!K_2I_{4,3}\lambda\!-\!K_1I_{4,4},\quad
{\sf R}\left(^{\lambda,\;{\bf x^4}}_{1,\;S_4}\right)\!=\!L_1\lambda^4\!+\!L_2I_{4,1}
\lambda^3\!+\!L_3I_{4,2}\lambda^2\!+\!L_2I_{4,3}\lambda\!+\!L_1I_{4,4},\nonumber\\
&&\hspace{-.8cm}
{\sf S}\left(^{\lambda,\;{\bf x^3}}_{2,\;S_3}\right)=
K_1\lambda^6\!+\!K_2I_{3,1}\lambda^5\!+\!\left(K_3I_{3,1}^2\!+\!K_4I_{3,2}\right)
\lambda^4\!-\!\left(K_3I_{3,2}^2\!+\!K_4I_{3,1}I_{3,3}\right)\lambda^2\!-\!K_2I_{3,2}I_{3,3}
\lambda\!-\!K_1I_{3,3}^2,\nonumber\\
&&\hspace{-.8cm}
{\sf R}\left(^{\lambda,\;{\bf x^3}}_{2,\;S_3}\right)=
L_1\lambda^6+L_2I_{3,1}\lambda^5+\left(L_3I_{3,1}^2+L_4I_{3,2}
\right)\lambda^4+(L_5I_{3,1}I_{3,2}+L_6I_{3,3})\lambda^3+
\nonumber\\
&&\hspace{1cm}\left(L_3I_{3,2}^2+L_4I_{3,1}I_{3,3}\right)\lambda^2+
L_2I_{3,2}I_{3,3}\lambda+L_1I_{3,3}^2,\nonumber\\
&&\hspace{-.8cm}
{\sf S}\left(^{\lambda,\;{\bf x^4}}_{2,\;S_4}\right)=K_1\lambda^8+
K_2I_{4,1}\lambda^7+\left(K_3I_{4,1}^2+K_4I_{4,2}\right)\lambda^6+\left(
K_5I_{4,1}I_{4,2}+K_6I_{4,3}\right)\lambda^5-\nonumber\\
&&\hspace{1cm}\left(K_5I_{4,3}I_{4,2}+K_6I_{4,1}I_{4,4}\right)\lambda^3-
\left(K_3I_{4,3}^2+K_4I_{4,2}I_{4,4}\right)\lambda^2-
K_2I_{4,3}I_{4,4}\lambda-K_1I_{4,4}^2,\nonumber\\
&&\hspace{-.8cm}
{\sf R}\left(^{\lambda,\;{\bf x^4}}_{2,\;S_4}\right)=L_1\lambda^8+
L_2I_{4,1}\lambda^7+\left(L_3I_{4,1}^2+L_4I_{4,2}\right)\lambda^6+
\left(L_5I_{4,1}I_{4,2}+L_6I_{4,3}\right)\lambda^5+
\left(L_7I_{4,2}^2+L_8I_{4,1}I_{4,3}+\right.\nonumber\\
&&\hspace{1cm}\left.L_9I_{4,4}\right)\lambda^4+
\left(L_5I_{4,3}I_{4,2}+L_6I_{4,1}I_{4,4}\right)\lambda^3+
\left(L_3I_{4,3}^2+L_4I_{4,2}I_{4,4}\right)\lambda^2+
L_2I_{4,3}I_{4,4}\lambda+L_1I_{4,4}^2.\nonumber
\eea}
The cubic polynomial (\ref{a3}) coincides with the first admitted self-dual polynomial ${\sf S}\left(^{\lambda,\;{\bf x^3}}_{1,\;S_3
}\right)$ in (\ref{d12}).
%%%%%%%%%%%%%%%%%%%%%%%%%%%%%%%%%%%%%%%%%%%
%%%%%%%%%%%%%%%%%%%%%%%%%%%%%%%%%%%%%%%%%%%
\subsection{Unimodality numbers and factorization of polynomials 
${\sf S}\left(^{\lambda,\;{\bf x^3}}_{2,\;S_3}\right)$ and 
${\sf S}\left(^{\lambda,\;{\bf x^4}}_{2,\;S_4}\right)$}\label{part41}
%%%%%%%%%%%%%%%%%%%%%%%%%%%%%%%%%%%%%%%%%%%
%%%%%%%%%%%%%%%%%%%%%%%%%%%%%%%%%%%%%%%%%%%
The numbers of independent coefficients $L_j$ and $K_j$ in (\ref{d12}) are called \cite{fel22} the {\it unimodality numbers} of self-dual polynomials and are denoted by $\mu\left\{{\sf R}\left(^{\lambda,\;{\bf x^n}}_{m,\;S_n}\right)\right\}$ and $\mu\left\{{\sf S}\left(^{\lambda,\;{\bf x^n}}_{m,\;S_n}\right)\right\}$, respectively. General expressions for the unimodality numbers were derived in \cite{fel22} using restriction partition functions with constraints. Their direct calculation for the polynomials in (\ref{d12}) yields, 
\bea
&&\hspace{-.5cm}
\mu\left[{\sf S}\left(^{\lambda,\;{\bf x^2}}_{1,\;S_2}\right)\right]\!=\!0,\quad
\mu\left[{\sf R}\left(^{\lambda,\;{\bf x^2}}_{1,\;S_2}\right)\right]\!=\!2,\quad
\mu\left[{\sf S}\left(^{\lambda,\;{\bf x^2}}_{2,\;S_2}\right)\right]\!=\!2,\quad\;
\mu\left[{\sf R}\left(^{\lambda,\;{\bf x^2}}_{2,\;S_2}\right)\right]\!=\!4,,\nonumber\\
&&\hspace{-.5cm}
\mu\left[{\sf S}\left(^{\lambda,\;{\bf x^3}}_{1,\;S_3}\right)\right]\!=\!2,\quad
\mu\left[{\sf R}\left(^{\lambda,\;{\bf x^3}}_{1,\;S_3}\right)\right]\!=\!2,\quad
\mu\left[{\sf S}\left(^{\lambda,\;{\bf x^3}}_{2,\;S_3}\right)\right]\!=\!4,\quad\;
\mu\left[{\sf R}\left(^{\lambda,\;{\bf x^3}}_{2,\;S_3}\right)\right]\!=\!6,,\nonumber\\
&&\hspace{-.5cm}
\mu\left[{\sf S}\left(^{\lambda,\;{\bf x^4}}_{1,\;S_4}\right)\right]\!=\!2,\quad
\mu\left[{\sf R}\left(^{\lambda,\;{\bf x^4}}_{1,\;S_4}\right)\right]\!=\!3,\quad
\mu\left[{\sf S}\left(^{\lambda,\;{\bf x^4}}_{2,\;S_4}\right)\right]\!=\!6,\quad\;
\mu\left[{\sf R}\left(^{\lambda,\;{\bf x^4}}_{2,\;S_4}\right)\right]\!=\!9.\label{d13}
\eea

%%%%%%%%%%%%%%%%%%%%%%%%%%%%%%%%%%%%%%%%%%%
Let ${\mathbb R}\left(^{\lambda,}_{m,\;S_n}\right)$ and ${\mathbb
 S}\left(^{\lambda,}_{m,\;S_n}\right)$ denote two sets of polynomials ${\sf R}\left(^{\lambda,\;\;{\bf x^n}}_{m_1,S_n}\right)$ and ${\sf S}\left(^{\lambda,\;\;{\bf x^n}}_{m_1,S_n}\right)$, respectively, with fixed $n$ and $m$ and varying sets of parameters ${\cal R}^{s,\;l}_{m,n}$ and ${\cal S}^{s,\;l}_{m,n}$, defined in (\ref{d6}, \ref{d7}). The following containments then hold (see \cite{fel22}, Lemma 4) 
\bea
&&{\sf R}\left(^{\lambda,\;\;{\bf x^n}}_{m_1,S_n}\right){\sf R}\left(^{\lambda,\;\;{\bf x^n}}_{m_2,S_n}\right),\;
{\sf S}\left(^{\lambda,\;\;{\bf x^n}}_{m_1,S_n}\right)
{\sf S}\left(^{\lambda,\;\;{\bf x^n}}_{m_2,S_n}\right)\in{\mathbb R}\left(^{\lambda,}_{m,S_n}\right),\quad m=m_1+m_2,\nonumber\\
&&{\sf R}\left(^{\lambda,\;\;{\bf x^n}}_{m_1,S_n}\right){\sf S}\left(^{\lambda,\;\;{\bf x^n}}_{m_2,S_n}\right),\;\;{\sf S}
\left(^{\lambda,\;\;{\bf x^n}}_{m_1,S_n}\right)
{\sf R}\left(^{\lambda,\;\;{\bf x^n}}_{m_2,S_n}\right)\in
{\mathbb S}\left(^{\lambda,}_{m,S_n}\right).\label{d14}
\eea
Following \cite{fel22}, section 4, we denote the differences
\bea
&&\hspace{-1cm}
{\cal M}_{RR}^R\left(^{\lambda,\;{\bf x^n}}_{m,S_n}\right)=
\mu\left[{\sf R}\left(^{\lambda,\;{\bf x^n}}_{m,S_n}\right)\right]-\mu\left[{\sf R}
\left(^{\lambda,\;\;{\bf x^n}}_{m_1,S_n}\right)\right]-\mu\left[{\sf R}\left(^{\lambda,
\;\;{\bf x^n}}_{m_2,S_n}\right)\right],\quad m=m_1+m_2,\nonumber\\
&&\hspace{-1cm}
{\cal M}_{SS}^R\left(^{\lambda,\;{\bf x^n}}_{m,S_n}\right)=
\mu\left[{\sf R}\left(^{\lambda,\;{\bf x^n}}_{m,S_n}\right)\right]-\mu\left[{\sf S}
\left(^{\lambda,\;\;{\bf x^n}}_{m_1,S_n}\right)\right]-\mu\left[{\sf S}\left(^{\lambda,
\;\;{\bf x^n}}_{m_2,S_n}\right)\right],\nonumber\\
&&\hspace{-1cm}
{\cal M}_{RS}^S\left(^{\lambda,\;{\bf x^n}}_{m,S_n}\right)=
\mu\left[{\sf R}\left(^{\lambda,\;{\bf x^n}}_{m,S_n}\right)\right]-\mu\left[{\sf R}
\left(^{\lambda,\;\;{\bf x^n}}_{m_1,S_n}\right)\right]-\mu\left[{\sf S}\left(^{\lambda,
\;\;{\bf x^n}}_{m_2,S_n}\right)\right].\label{d15}
\eea
The necessary conditions for containment (\ref{d14}) are
\bea
a)\quad {\cal M}_{RR}^R\left(^{\lambda,\;{\bf x^n}}_{m,S_n}\right)\ge 0,\qquad b)\quad {\cal 
M}_{SS}^R\left(^{\lambda,\;{\bf x^n}}_{m,S_n}\right)\ge 0,\qquad c)\quad {\cal 
M}_{RS}^S\left(^{\lambda,\;{\bf x^n}}_{m,S_n}\right)\ge 0.\label{d16}
\eea

Consider a factorization of polynomials ${\sf S}\left(^{\lambda,\;{\bf x^n}}_{m,\;S_n}\right)$. According to Lemma 6 in \cite{fel22}, the identity
\bea
{\sf S}\left(^{\lambda,\;{\bf x^2}}_{m,\;S_2}\right)={\sf S}\left(^{\lambda,\;{\bf x^2}}_{1,\;S_2}\right){\sf R}
\left(^{\lambda,\quad\;{\bf x^2}}_{m-1,\;S_2}\right)\label{d17}
\eea
holds irrespective to $m\ge 1$. Bearing in mind inequalities 
${\cal R}^{s,\;l}_{m,n},\;{\cal S}^{s,\;l}_{m,n}\ge 0$ from (\ref{d8}), we conclude that $\lambda({\bf x}^2)=\sqrt{x_1x_2}\;$ is a unique positive root of every polynomial ${\sf S}\left(^{
\lambda,\;{\bf x^2}}_{m,\;S_2}\right)$.

\begin{figure}[h]\begin{center}
\psfig{figure=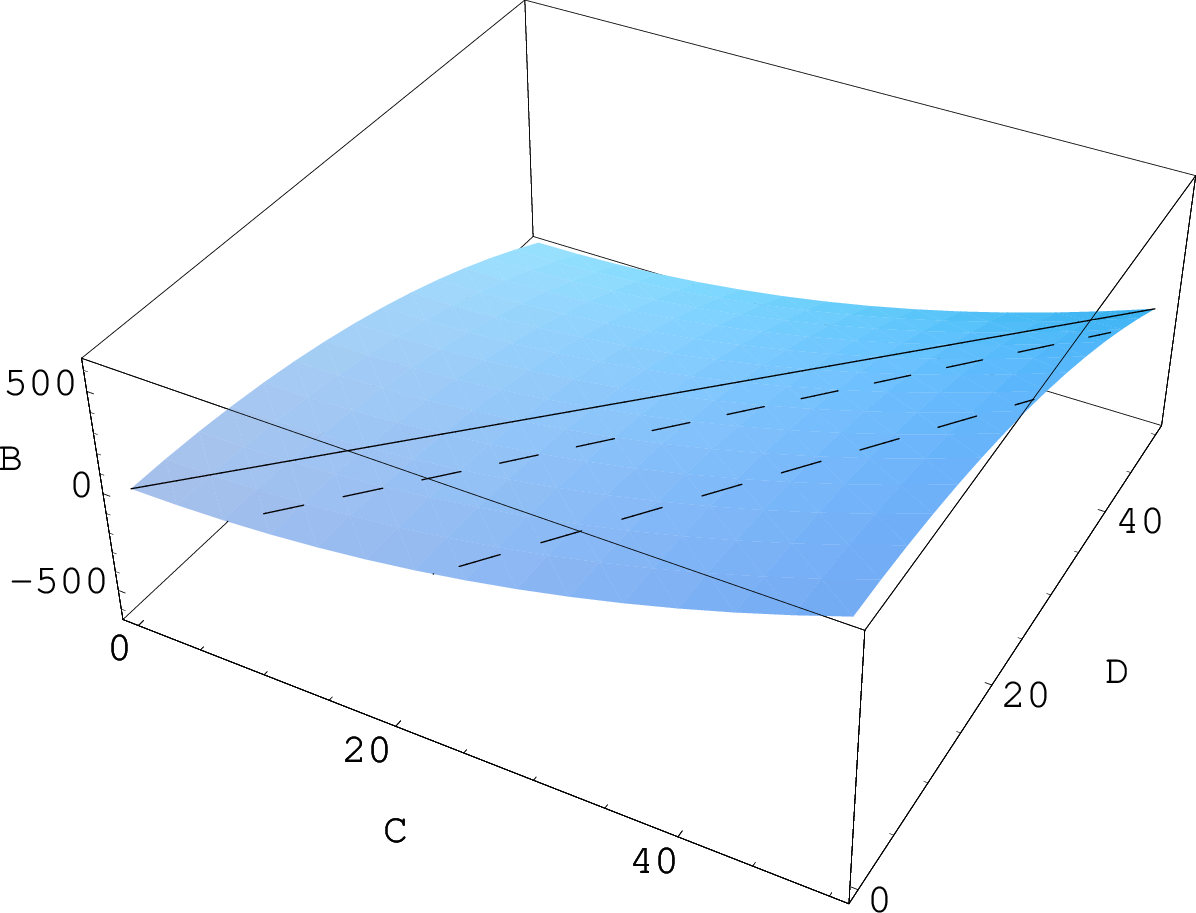,height=6cm}
\end{center}
\caption{A part of the hyperbolic surface ${\cal H}^2$, where solutions of equations ${\sf S}\left(^{\lambda,\;{\bf x^3}}_{1,\;S_
3}\right)\!=\!0$ and ${\sf S}\left(^{\lambda,\;{\bf x^3}}_{2,\;S_3}
\right)\!=\!0$ are coincided in the Eucleadian space ${\mathbb E}^3
\!=\!\left\{x_1,x_2,x_3\right\}$. Three lines ${\bf \Bbb{R}}^1(S_3)$, lying on the hyperbolic surface ${\cal H}^2$, correspond to three distinct values of $A\!=\!(D-C)/(2Q_1)$, concerned with three  2D three-color regular structures in Figure \ref{fig1}, $A_{Rh}\!=0,$ $A_{Fl}\!=3.76,$ $A_{He}\!=11.37$, and the cubic equation (\ref{a3}).}\label{fig4}
\end{figure}

The uniqueness of the positive root $\lambda({\bf x}^n)$ of the polynomials ${\sf S}\left(^{\lambda,\;{\bf x^n}}_{m,\;S_n}\right)$ vanishes when passing to $n\ge 3$. To demonstrate this, we factorize the polynomial ${\sf S}\left(^{\lambda,\;{\bf x^3}}_{2,\;S_3}\right)$ and, according to (\ref{d13},\ref{d14}, \ref{d16} c), we obtain
\bea
{\sf S}\left(^{\lambda,\;{\bf x^3}}_{2,\;S_3}\right)={\sf S}\left(^{\lambda,\;{\bf x^3}}_{1,\;S_3}\right){\sf R}\left(^
{\lambda,\;{\bf x^3}}_{1,\;S_3}\right).\label{d18}
\eea
We substitute the expressions of the self-dual polynomials (\ref{d12}) into (\ref{d18}), and, for greater clarity, replace all coefficients $K_j$ by $Q_j$ in the polynomial ${\sf S}\left(^{
\lambda,\;{\bf x^3}}_{2,\;S_3}\right)$ in (\ref{d12}),
\bea
&&\hspace{-.8cm}
\left(K_1\lambda^3+K_2 I_{3,1}\lambda^2-K_2I_{3,2}\lambda- K_1I_{3,3}\right)\times
\left(L_1\lambda^3+L_2I_{3,1}\lambda^2+L_2I_{3,2}\lambda+L_1I_{3,3}\right)=\nonumber\\
&&\hspace{-.8cm}
Q_1\lambda^6+Q_2I_{3,1}\lambda^5+\left(Q_3I_{3,1}^2+Q_4I_{3,2}\right)\lambda^4-
\left(Q_3I_{3,2}^2+Q_4I_{3,1}I_{3,3}\right)\lambda^2-Q_2 I_{3,2}I_{3,3}\lambda-Q_1I_{3,3}^2.
\nonumber
\eea
Equating degrees of $\lambda$ in both sides of the last identity
\bea
Q_1=K_1L_1,\quad Q_2=K_1L_2+K_2L_1,\quad Q_3=K_2L_2,\quad Q_4=K_1L_2-K_2L_1,\label{d19}
\eea
we obtain a quadric 3-surface ${\cal H}_1(S_3)$ embedded in the 4D Eucleadian space ${\mathbb Q}^4(S_3)=\left\{Q_1,\ldots,Q_4\right\}$
\bea
Q_2^2-Q_4^2=4Q_1Q_3.\label{d20}
\eea
To visualize this embedding, in Figure \ref{fig4} we show the hyperbolic surface ${\cal H}^2:D^2-C^2=4B$, embedded in the 3D Eucleadian space ${\mathbb E}^3\!=\!\left\{B,C,D\right\}$ with the
renamed variables, $B=Q_1Q_3$, $C=Q_4$, $D\!=\!Q_2$.
%%%%%%%%%%%%%%%%%%%%%%%%%%%%%%%%%%%%%%%%%%%
%%%%%%%%%%%%%%%%%%%%%%%%%%%%%%%%%%%%%%%%%%%

Similar reasoning in the case $n=4$ shows that a factorization
\bea
{\sf S}\left(^{\lambda,\;{\bf x^4}}_{2,\;S_4}\right)={\sf S}\left(^{\lambda,\;{\bf x^4}}_{1,\;S_4}\right){\sf R}\left(^
{\lambda,\;{\bf x^4}}_{1,\;S_4}\right)\label{d21}
\eea
holds if and only if the following equalities for the coefficients $K_j$, which are replaced by $Q_j$ in the polynomial ${\sf S}\left(^{\lambda,\;{\bf x^4}}_{2,\;S_4}\right)$ in (\ref{d12}), are satisfied
\bea
Q_1\!=\!K_1L_1,\quad Q_2\!=\!K_1L_2+K_2L_1,\quad Q_3\!=\!K_2L_2,\quad Q_4\!=\!K_1L_3,\quad Q_5\!=\!K_2L_3,\quad 
Q_6\!=\!K_1L_2-K_2L_1.\nonumber
\eea
Six relations for five variables lead to two independent identities
\bea
a)\quad Q_2^2-Q_6^2=4Q_1Q_3,\qquad b)\quad \left(\frac{Q_4}{Q_5}\right)^2=\frac{Q_1}{Q_3}\frac{Q_2+Q_6}{Q_2-Q_5}.\label{d22}
\eea
These identities define in the 6D Eucleadian space ${\mathbb Q}^6(S_4)=\left\{Q_1,\ldots,Q_6\right\}$ two manifolds, a quadric 3-surface ${\cal H}_1(S_4)$ and a quartic 5-surface ${\cal H}_2(S_4)$. The unique positive solutions of equations ${\sf S}\left(^{\lambda,\;{\bf x^4}
}_{1,\;S_4}\right)\!=\!0$ and ${\sf S}\left(^{\lambda,\;{\bf x^4}}_
{2,\;S_4}\right)\!=\!0$ coincide in the entire Euclidean space 
${\mathbb E}^4=\left\{x_1,\ldots,x_4\right\}$ if the coefficients $Q_j$ of the polynomial ${\sf S}\left(^{\lambda,\;{\bf x^4}}_{2,\;
S_4}\right)$ lie at the intersection of the two hypersurfaces 
${\cal H}_1(S_4)\cap{\cal H}_2(S_4)$.

If (\ref{d16} c) holds, then a factorization can be continued, obtaining a sequence of embedded spaces 
\bea
&&\hspace{-1cm}
{\bf \Bbb{Q}}^2(S_3)\;\subset\;{\cal L}_1(S_3)\;\subset\;{\bf \Bbb{Q}}^4(S_3)\;\subset\;{\cal L}_2(S_3)\;\subset\;{\bf \Bbb{Q}}^
{10}(S_3)\;\subset\;\ldots\;\subset\;{\bf\Bbb{Q}}^{\mu\left[{\sf 
S}\left(^{\lambda,\;{\bf x^3}}_{m,S_3}\right)\right]},\label{d23}\\
&&\hspace{-1cm}
{\bf \Bbb{Q}}^2(S_4)\;\subset\;{\cal L}_1(S_4)\;\subset\;{\bf \Bbb{Q}}^6(S_4)\;\subset\;{\cal L}_2(S_4)\;\subset\;{\bf \Bbb{Q}}^
{16}(S_4)\;\subset\;\ldots\;\subset\;{\bf\Bbb{Q}}^{\mu\left[{\sf S}
\left(^{\lambda,\;{\bf x^4}}_{m,S_4}\right)\right]},\nonumber
\eea
where ${\bf\Bbb{Q}}^{\mu\left[{\sf S}\left(^{\lambda,\;{\bf x^n}}_
{m,S_n}\right)\right]}$ denotes the Eucleadian space spanned on the $\mu\left[{\sf S}\left(^{\lambda,\;{\bf x^n}}_{m,S_n}\right)
\right]$ polynomial coefficients as on unit vectors. A manifold 
${\cal L}_m(S_n)$ is defined by a factorization of a self-dual polynomial ${\sf S}\left(^{\lambda,\;{\bf x^n}}_{m,\;S_n}\right)$
 and is given by the intersection of $\nu_m(S_n)=1+{\cal M}_{RS}^S
\left(^{\lambda,\quad\;{\bf x^n}}_{m+1,S_n}\right)$ hypersurfaces
${\cal H}_j(S_n)$,
\bea
{\cal L}_m(S_n)=\bigcap_{j=1}^{\nu_m(S_n)}{\cal H}_j(S_n),\qquad
\mbox{e.g.,}\quad 
{\cal L}_1(S_3)={\cal H}_1(S_3),\quad {\cal L}_1(S_4)={\cal H}_1
(S_4)\cap{\cal H}_2(S_4).\nonumber
\eea

The sequence (\ref{d23}) of embedded parametric spaces paves the way for successive approximation of numerical calculations in the set of self-dual polynomials ${\sf S}\left(^{\lambda,\;\;{\bf x^n}}_{m,\;S_n}\right)$ with increasing $m$. In other words, the accuracy of fitting the cubic equation ${\sf S}\left(^{\lambda,\;\;{\bf x^3}}_{1,\;\;S_3}\right)\!=\!0$ in (\ref{a3}) to numerical calculations \cite{fel00} for four different checkerboards (see Figure \ref{fig1}) can be enhanced by passing to self-dual equations ${\sf S}\left(^{\lambda,\;\;{\bf x^3}}_{m,\;S_3}\right)=
0$ with higher $m\ge 2$. For example, a new equation ${\sf S}\left(^{\lambda,\;\;{\bf x^3}}_{2,\;\;S_3}\right)\!=\!0$ provides a better fit, possessing coefficients $Q_j$, whose image $(Q_1,Q_2,
Q_3,Q_4)\in{\bf \Bbb{Q}}^4(S_3)$ is in the vicinity of ${\cal H}_1(S_3)$, but not in ${\cal H}_1(S_3)$.
%%%%%%%%%%%%%%%%%%%%%%%%%%%%%%%%%%%%%%%%%%%
%%%%%%%%%%%%%%%%%%%%%%%%%%%%%%%%%%%%%%%%%%%
\section{Self-dual polynomial ${\sf S}\left(^{\lambda,\;\;{\bf x^n}}_{m,\;S_n}\right)$ and its root $\lambda({\bf x}^n)$}\label{part5}
%%%%%%%%%%%%%%%%%%%%%%%%%%%%%%%%%%%%%%%%%%%
%%%%%%%%%%%%%%%%%%%%%%%%%%%%%%%%%%%%%%%%%%%
\noindent
Let $\lambda({\bf x}^n)$ be a real solution to the equation ${\sf S}\left(^{\lambda,\;\;{\bf x^n}}_{m,\;S_n}\right)=0$. Following \cite{fel22}, Section 5, we recall important algebraic properties of this solution.
%%%%%%%%%%%%%%%%%%%%%%%%%%%%%%%%%%%%%%%%%%%

$\bullet$
{\sf Homogeneity of 1st order and permutation invariance}
\bea
\lambda(kx_1,\ldots,kx_n)=k\;\lambda(x_1,\ldots,x_n),\qquad
\lambda({\widehat {\cal P}}\;\{x_1,\ldots,x_n\})=\;\lambda(x_1,
\ldots,x_n),\label{e1}
\eea
where ${\widehat {\cal P}}\in S_n$ is a permutation operator of the indices $\{1,\ldots,n\}$.

%%%%%%%%%%%%%%%%%%%%%%%%%%%%%%%%%%%%%%%%%%%
$\bullet$
{\sf Self-duality}

\noindent
An invariance of $\lambda({\bf x}^n)$ under the action (\ref{d11}) of inversion ${\sf Q}:\{x_i\rightarrow x_i^{-1},\;\lambda
\rightarrow\lambda^{-1}\}$ results in
\bea
\lambda(x_1,\ldots,x_n)\times\lambda\left(x_1^{-1},\ldots,x_n^{-1}\right)=1.\label{e2}
\eea

%%%%%%%%%%%%%%%%%%%%%%%%%%%%%%%%%%%%%%%%%%%
$\bullet$
{\sf Compatibility of $\lambda({\bf x}^n)$}
\bea
\lambda(x,\ldots,x)=x.\label{e3}
\eea

%%%%%%%%%%%%%%%%%%%%%%%%%%%%%%%%%%%%%%%%%%%
$\bullet$
{\sf Positivity and Monotony}
\bea
\lambda\left({\bf x}^n\right)\ge 0,\qquad \partial\lambda/\partial x_i>0,\quad 1\le i\le n,\qquad {\bf x^n}\in{\mathbb E}^n_>.\label{e4}
\eea

%%%%%%%%%%%%%%%%%%%%%%%%%%%%%%%%%%%%%%%%%%%
$\bullet$
{\sf Uniqueness of positive solution $\lambda({\bf x}^n)$ }

\noindent
By Descartes' rule of signs \cite{dic22}, the polynomial ${\sf S}\left(^{\lambda,{\bf x^n}}_{m,S_n}\right)$ has a unique positive root.

%%%%%%%%%%%%%%%%%%%%%%%%%%%%%%%%%%%%%%%%%%%
$\bullet$
{\sf Common solution $\lambda({\bf x}^n)$ of equations ${\sf S}\left(^{\lambda,\;\;{\bf x^n}}_{m_2,S_n}\right)=0$}

\noindent
If there exists such a $\lambda_{*}$, that for each $x_i$ it is possible to determine $x_{j}$, such that
\bea
x_i \cdot x_{j}=\lambda_*^2\qquad\mbox{then}\qquad\lambda\left(
{\bf x^n}\right)=\lambda_*,\label{e5} 
\eea
and $\lambda_*$ is independent of $m$. The next statement is equivalent to the previous one \cite{fel22}: 
\bea
\mbox{if}\qquad I_{n,r}^n\;I_{n,n}^{n-2r}=I_{n,n-r}^n,\;\;r=1,\ldots,\left\lfloor\frac{n}{2}\right\rfloor\qquad\mbox{then}\qquad
\lambda\left({\bf x^n}\right)=\sqrt[n]{I_{n,n}}.\label{e6}
\eea
For $n=1,2$ this equivalence is obvious; for $n=3,4$ it can be shown by factorization of invariants (\ref{e6})
\bea
&&\hspace{-1cm}
n=3\;:\;\;I_{3,3} I_{3,1}^3=I_{3,2}^3\quad\longrightarrow\quad
(x_1^2-x_2 x_3)(x_2^2-x_3 x_1)(x_3^2-x_1x_2)=0,\label{e7}\\
&&\hspace{-1cm}
n=4\;:\;\;I_{4,4} I_{4,1}^2=I_{4,3}^2\quad\longrightarrow\quad
(x_1 x_2- x_3 x_4)(x_1 x_3- x_2 x_4)(x_1 x_4-x_2 x_3)=0.\label{e8}
\eea

%%%%%%%%%%%%%%%%%%%%%%%%%%%%%%%%%%%%%%%%%%%
$\bullet$
{\sf Bruggeman's polynomials $Br(\lambda,{\bf x}^n)$}

\noindent
A special case of self-dual polynomials ${\sf S}\left(^{\lambda,\;{\bf x^{2n+1}}}_{1,\;S_{2n+1}}\right)$ and ${\sf S}\left(^{\lambda,\;{\bf x^{2n}}}_{1,\;S_{2n}}\right)$ is given by
\bea
Br(\lambda,{\bf x}^{2n+1})=\lambda^{2n+1}+\frac{2n-1}{2n+1}I_{2n+1,1}\lambda^{2n}+
\frac{2n-3}{2n+1}I_{2n+1,2}\lambda^{2n-1}+\ldots+\frac1{2n+1}I_{2n+1,n-1}\lambda^{n+1}-\nonumber\\
\frac1{2n+1}I_{2n+1,n+1}\lambda^{n-1}-\ldots-\frac{2n-3}{2n+1}I_{2n+1,2n-1}\lambda^2-\frac{2n-1}{2n+1}I_{2n+1,2n}\lambda-I_{2n+1,2n+1},\nonumber
\eea
\bea
\hspace{-2cm}
Br(\lambda,{\bf x}^{2n})=\lambda^{2n}+\frac{n-1}{n}I_{2n,1}\lambda^{2n-1}+\frac{n-2}{n}I_{2n,2}\lambda^{2n-2}+\ldots+
\frac1{n}I_{2n,n-1}\lambda^{n+1}-\nonumber\\
\hspace{-2cm}
\frac1{n}I_{2n,n+1}\lambda^{n-1}-\ldots-\frac{n-2}{n}I_{2n,2n-2}\lambda^2-\frac{n-1}{n}I_{2n,2n-1}\lambda-I_{2n,2n}.\label{e9}
\eea

We emphasize a remarkable coincidence between the basic properties of the solutions $\lambda\left({\bf x^n}\right)$ of the self-dual equations ${\sf S}\left(^{\lambda,\;{\bf x^n}}_{m,\;S_n}\right)=0$
in Section \ref{part5} and the effective conductivity $\sigma_e(
\sigma_1,\ldots,\sigma_n)$ in Section \ref{part3}. The following correspondence between these two sets of formulas
\bea
&&\mbox{(\ref{e1})}\;\;\longleftrightarrow\;\;\mbox{(\ref{c3}, \ref{c4})},\qquad
\mbox{(\ref{e2})}\;\;\longleftrightarrow\;\;\mbox{(\ref{c5})},\qquad
\mbox{(\ref{e3})}\;\;\longleftrightarrow\;\;\mbox{(\ref{c6})},\nonumber\\
&&\mbox{(\ref{e4})}\;\;\longleftrightarrow\;\;\mbox{(\ref{c7})},\qquad
\mbox{(\ref{e5}, \ref{e6})}\;\;\longleftrightarrow\;\;\mbox{(\ref{c8})},\qquad
\mbox{(\ref{e9})}\;\;\longleftrightarrow\;\;\mbox{(\ref{c9})},\label{e10}
\eea
strongly suggests
\footnote{By the correspondence $\mbox{(\ref{e9})}\longleftrightarrow\mbox{(\ref{c9})}$ in (\ref{e10}) we mean that self-dual equations $Br(\lambda,{\bf x}^{2n})=0$ and $Br(\lambda,{\bf x}^{2n+1})=0$ can be represented as the Bruggeman equation $\sum_{j=1}^m(\lambda-x_j)/(\lambda+x_j)=0$, where $m=2n$ or $m=2n+1$, respectively.}
properties such as permutation invariance, self-duality, compatibility, positivity, monotony and Dykhne's ansatz as a good basis for applying the algebra of self-dual symmetric polynomials to the isotropic effective conductivity problem in 2D composite.

This application has several important implications. One of them is an affirmative answer to the question: Can the self-dual equation 
${\sf S}\left(^{\lambda,\;{\bf x^n}}_{m,\;S_n}\right)=0$ with $m\geq 2$ improve the precision $\left|\epsilon_{max}^1\right|$ of the first approximation, which is achieved in equation ${\sf S}\left(^{\lambda,\;{\bf x^n}}_{1,\;S_n}\right)=0$, for all structures studied numerically in \cite{fel00}, that is, $\left|\epsilon_{max}^m\right|\leq \left|\epsilon_{max}^1\right|$ (see Section \ref{part41} of the present paper).

Another important conclusion concerns the upper and lower universal bounds of the roots $\lambda({\bf x}^n)$ of self-dual polynomials 
${\sf S}\left(^{\lambda,\;{\bf x^n}}_{m,\;S_n}\right)$, $n=3,4$, which we discuss in Sections \ref{part51}, \ref{part52}, and  \ref{part53}.
%%%%%%%%%%%%%%%%%%%%%%%%%%%%%%%%%%%%%%%%%%%
%%%%%%%%%%%%%%%%%%%%%%%%%%%%%%%%%%%%%%%%%%%
\subsection{Proper self-dual polynomials ${\sf S}_{\odot}\left(^{
\lambda,\;{\bf x^n}}_{m,\;S_n}\right)$}\label{part51}
%%%%%%%%%%%%%%%%%%%%%%%%%%%%%%%%%%%%%%%%%%%
%%%%%%%%%%%%%%%%%%%%%%%%%%%%%%%%%%%%%%%%%%%
To simplify a study of the upper and lower bounds on the positive roots $\lambda({\bf x}^n)$ of self-dual polynomials and to adapt the theory to physical purposes, we set out two more requirements.

We can represent the expression for ${\sf S}\left(^{\lambda,\;
{\bf x^n}}_{m,\;S_n}\right)$ in (\ref{d5}) as follows
\bea
&&\hspace{-1.5cm}
{\sf S}\left(^{\lambda,\;{\bf x^n}}_{m,S_n}\right)=\sum_{s=0}^{\left\lfloor 
s_*\right\rfloor}\sum_{l=1}^{P_n^m(s)}{\cal S}^{s,\;l}_{m,n}\;\Theta^{s,l}
\left(^{\lambda,\;{\bf x^n}}_{m,S_n}\right),\quad {\cal S}^{s,l}_{m,n}\ge 0,
\qquad \alpha_r=\alpha_r\left(^{s,\;\;l}_{m,S_n}\right),\label{e11}\\
&&\hspace{-1.5cm}
\Theta^{s,l}\left(^{\lambda,\;{\bf x^n}}_{m,S_n}\right)=\lambda^s\prod_{r=0}^n
I_{n,r}^{\alpha_r}\left[\lambda^{mn-2s}-\Gamma\left(^{s,\;\;l}_{m,S_n}\right)
\right],\qquad\Gamma\left(^{s,\;\;l}_{m,S_n}\right)=\prod_{r=0}^n
\left(\frac{I_{n,n-r}}{I_{n,r}}\right)^{\alpha_r}.\nonumber
\eea
If $mn=0\;(\bmod\;2)$, i.e., $s_*=mn/2$ is an integer, then the middle term in the summands (\ref{e11}) reads
\bea
\Theta^{s_*,l}\left(^{\lambda,\;{\bf x^n}}_{m,S_n}\right)=
\lambda^{mn/2}\left(\;\prod_{r=0}^nI_{n,r}^{\alpha_r}-\prod_
{r=0}^nI_{n,n-r}^{\alpha_r}\right),\label{e12}
\eea
where $\alpha_r$ satisfy two Diophantine equations (\ref{d9}). In many cases both products in (\ref{e12}) annihilate and the entire term does not contribute to ${\sf S}\left(^{\lambda,\;{\bf x^n}}_
{m,S_n}\right)$. That is true for all self-dual polynomials with $m=1,2$, but it appears already in ${\sf S}\left(^{\lambda,\;
{\bf x^4}}_{3,\;S_4}\right)$, that is, $\lambda^6\left(I_{4,1}^2
I_{4,4}-I_{4,3}^2\right)$ (see \cite{fel22}, Appendix A).

The proof of the existence of upper and lower bounds (see \cite{fel22}, Section 6) on the positive root $\lambda({\bf x}^n)$ of self-dual polynomials ${\sf S}\left(^{\lambda,\;{\bf x^n}}_{m,S_n}\right)$ is essentially based on the absence of specific terms (\ref{e12}). Therefore, the first additional requirement reads
\bea
{\cal S}^{s_*,\;l}_{m,\;n}=0,\qquad\mbox{if}\qquad mn=0\;(\bmod\;2).\label{e13}
\eea

Another claim comes from physical considerations: the upper bound of $\lambda({\bf x}^n)$ should not diverge if some of the variables $x_j$ are vanishing. To illustrate this, consider $\Gamma\left(^{s,\;\;l}_{m,S_n}\right)$ in (\ref{e11}), 
\bea
\Gamma\left(^{s,\;\;l}_{m,S_n}\right)\simeq\epsilon^{\theta_k+2
\theta_{k-1}+\ldots+(k+1)\theta_0},\quad k\le\left\lfloor n/2
\right\rfloor,\qquad \theta_k=\alpha_k-\alpha_{n-k},\label{e14}
\eea
where $x_1,\ldots,x_k\simeq \epsilon\to 0$. Consequently, the necessary conditions for convergence of $\Gamma\left(^{s,\;\;l}_
{m,S_n}\right)$ in various combinations of vanishing $x_j$ read (see \cite{fel22}, inequalities 6.16)
\bea
\theta_0\ge 0,\quad \theta_1+2\theta_0\ge 0,\quad\ldots,\quad
\theta_k+2\theta_{k-1}+\ldots+(k+1)\theta_0\ge 0,\qquad
0\le k\le\left\lfloor n/2 \right\rfloor.\label{e15}
\eea
According to Theorem 6 in \cite{fel22}, if a set of inequalities (\ref{e15}) holds, then
\bea
\Gamma\left(^{s,\;\;l}_{m,S_n}\right)\le\left(\frac{I_{n,1}}{n}\right)^{mn-2s}.\label{e16}
\eea

If the self-dual polynomial ${\sf S}\left(^{\lambda,\;{\bf x^n}}_
{m,S_n}\right)$, defined in (\ref{e11}), satisfies the requirements of (\ref{e13}) and (\ref{e16}), we call it a {\em proper} polynomial and denote it by ${\sf S}_{\odot}\left(^{\lambda,\;
{\bf x^n}}_{m,S_n}\right)$.
%%%%%%%%%%%%%%%%%%%%%%%%%%%%%%%%%%%%%%%%%%%%%%%%%%%%%%%%%%%%%%%%%%
%%%%%%%%%%%%%%%%%%%%%%%%%%%%%%%%%%%%%%%%%%%%%%%%%%%%%%%%%%%%%%%%%%%
\subsection{Upper and lower bounds of $\lambda_{\odot}\left({\bf x}^3\right)$}\label{part52}
%%%%%%%%%%%%%%%%%%%%%%%%%%%%%%%%%%%%%%%%%%%%%%%%%%%%%%%%%%%%%%%%%%
%%%%%%%%%%%%%%%%%%%%%%%%%%%%%%%%%%%%%%%%%%%%%%%%%%%%%%%%%%%%%%%%%%
In this section, we provide {\em weak} universal bounds for the unique positive root $\lambda_{\odot}({\bf x}^n)$ of the proper self-dual polynomials ${\sf S}_{\odot}\left(^{\lambda,\;{\bf x^n}}_{m,S_n}\right)$ (see \cite{fel22}, Theorem 7),
\bea
\frac{nI_{n,n}}{I_{n,n-1}}\leq\lambda_{\odot}({\bf x}^n)\leq\;\frac{I_{n,1}}{n},\label{e17}
\eea
where the upper and lower bounds correspond to the arithmetic and harmonic meanss, respectively.

For small $n=3,4$, {\em strong} universal bounds can be found for the unique positive root $\lambda_{\odot}({\bf x}^n)$.

\noindent
If $n=3$, then following \cite{fel22}, Theorem 9, we get
\bea
\omega\left({\bf x^3}\right)\le\lambda_{\odot}\left({\bf x^3}\right)\le\Omega\left({\bf x^3}\right),\label{e18}
\eea
where the upper $\Omega\left({\bf x^3}\right)$ and lower $\omega
\left({\bf x^3}\right)$ bounds are given by
\bea
\Omega\left({\bf x^3}\right)=\max\left\{\frac{I_{3,2}}{I_{3,1}},\left(\frac{I_{3,1}I_{3,3}}{I_{3,2}}\right)^{1/2}\right\},\qquad 
\omega\left({\bf x^3}\right)=\min\left\{\frac{I_{3,2}}{I_{3,1}},\left(\frac{I_{3,1}I_{3,3}}{I_{3,2}}\right)^{1/2}\right\}.\label{e19}
\eea
The bounds in (\ref{e19}) are stronger than those in (\ref{e17}) (for the proof see in \cite{fel22}, Section 7.1),
\bea
\frac{3I_{3,3}}{I_{3,2}}\le\omega\left({\bf x^3}\right),\qquad
\Omega\left({\bf x^3}\right)\le\frac{I_{3,1}}{3}.\label{e20}
\eea
Decompose the entire Euclidean space ${\mathbb E}^3_{>}$ into three different subspaces
\bea
{\mathbb E}^3_>={\mathbb P}^3_+\cup{\mathbb P}^2_0\cup{\mathbb P}^
3_-\;,\qquad\left\{\begin{array}{lr}
{\mathbb P}^3_+=\left\{{\mathbb E}^3_>\;|\;I_{3,2}^3>I_{3,3}I_{3,1}^3\right\},&\dim{\mathbb P}^3_+=3,\\
{\mathbb P}^2_0\;=\left\{{\mathbb E}^3_>\;|\;I_{3,2}^3=I_{3,3}I_{3,1}^3\right\},&\dim{\mathbb P}^2_0\;=2,\\
{\mathbb P}^3_-=\left\{{\mathbb E}^3_>\;|\;I_{3,2}^3<I_{3,3}I_{3,1}^3\right\},
&\dim{\mathbb P}^3_-=3,\end{array}\right.\label{e21}
\eea
where ${\mathbb P}^2_0$ is a surface that separates ${\mathbb P}^3
_+$ and ${\mathbb P}^3_-$ and is given by equation (\ref{e7}). 

The formulas (\ref{e18}) and (\ref{e19}) are dual in the 
following sense:
\bea
\mbox{if}\quad\left(\frac{I_{3,1}I_{3,3}}{I_{3,2}}\right)^{1/2}\le
\lambda_{\odot}\left({\bf x^3}\right)\le\frac{I_{3,2}}{I_{3,1}},\quad {\bf x^3}\in{\mathbb P}^3_+,\nonumber\\
\mbox{then}\quad\frac{I_{3,2}}{I_{3,1}}\le\lambda_{\odot}\left({\bf x^3}\right)\le\left(\frac{I_{3,1}I_{3,3}}{I_{3,2}}\right)^{1/2},
\quad{\bf x^3}\in{\mathbb P}^3_-,\nonumber
\eea
and vice versa, the actual expressions will be exchanged for the lower and upper bounds in the subspaces ${\mathbb P}^3_+$ and 
${\mathbb P}^3_-$, while if ${\bf x^3}\in{\mathbb P}^2_0$, then 
$\Omega\left({\bf x^3}\right)=\omega\left({\bf x^3}\right)$.

Find the asymptotics of the upper and lower bounds,
\bea
x_1<<x_2,x_3,\;\left\{\begin{array}{l}
\Omega_3(x_i)\simeq x_2 x_3/(x_2+x_3)\\
\omega_3(x_i)\simeq\sqrt{x_1}\;\sqrt{x_2+x_3}\end{array}\right.,\;\;x_1>>x_2=x_3,\;\left\{ \begin{array}{l}\Omega_3(x_i)\simeq
\frac{1}{\sqrt{2}}\;\sqrt{x_1x_2}\\    
\omega_3(x_i)\simeq 2\;x_2\end{array}\right.\quad\label{e22}
\eea
and compare (\ref{e22}) with asymptotics (\ref{c11}) for the {\sf Rh} checkerboard \cite{kha00}. They match well in both subspaces 
${\mathbb P}^3_+$ and ${\mathbb P}^3_-$.

The formulas (\ref{e22}) give finite mutually dual bounds in two opposite cases
\bea
\Omega(0,x,x)=\frac{x}{2},\qquad\omega(\infty,x,x)=2x.\nonumber
\eea
These agree with numerical calculations of the effective conductivity of a three-phase {\sf He} composite \cite{fel00} in two special cases: $a$) one isolating and two other similar non-isolating phases and $b$) one superconducting and two other similar non-superconducting phases.
%%%%%%%%%%%%%%%%%%%%%%%%%%%%%%%%%%%%%%%%%%%%%%%%%%%%%%%%%%%%%%%%%%
%%%%%%%%%%%%%%%%%%%%%%%%%%%%%%%%%%%%%%%%%%%%%%%%%%%%%%%%%%%%%%%%%%%
\subsection{Upper and lower bounds of $\lambda_{\odot}({\bf x}^4)$}
\label{part53}
%%%%%%%%%%%%%%%%%%%%%%%%%%%%%%%%%%%%%%%%%%%%%%%%%%%%%%%%%%%%%%%%%%
%%%%%%%%%%%%%%%%%%%%%%%%%%%%%%%%%%%%%%%%%%%%%%%%%%%%%%%%%%%%%%%%%%
If $n=4$, then following \cite{fel22}, Theorem 10, we get
\bea
\omega\left({\bf x^4}\right)\le\lambda_{\odot}\left({\bf x^4}\right)\le\Omega\left({\bf x^4}\right),\label{e23}
\eea
where the upper $\Omega\left({\bf x^4}\right)$ and lower $\omega
\left({\bf x^4}\right)$ bounds are given by
\bea
\hspace{-.3cm}
\Omega\left({\bf x^4}\right)\!=\!\max\left\{\left(\frac{I_{4,3}}{I_{4,1}}\right)
^{1/2}\!\!,\left(\frac{I_{4,1}I_{4,4}}{I_{4,3}}\right)^{1/2}\right\},\quad
\omega\left({\bf x^4}\right)\!=\!\min\left\{\left(\frac{I_{4,3}}{I_{4,1}}\right)
^{1/2}\!\!,\left(\frac{I_{4,1}I_{4,4}}{I_{4,3}}\right)^{1/2}\right\}.\label{e24}
\eea
The bounds in (\ref{e24}) are stronger than those in (\ref{e17}), (for the proof see in \cite{fel22}, Section 7.2),
\bea
\frac{4I_{4,4}}{I_{4,3}}\le\omega\left({\bf x^4}\right),\qquad
\Omega\left({\bf x^4}\right)\le\frac{I_{4,1}}{4}.\label{e25} 
\eea
Decompose the entire Euclidean space ${\mathbb E}^4_{>}$ in three different subspaces
\bea
{\mathbb E}^4_>={\mathbb P}^4_+\cup{\mathbb P}^3_0\cup{\mathbb P}^4_-\;,\qquad
\left\{\begin{array}{lr}
{\mathbb P}^4_+=\left\{{\mathbb E}^4_>\;|\;I_{4,3}^2>I_{4,4}I_{4,1}^2\right\},&
\dim{\mathbb P}^4_+=4,\\
{\mathbb P}^3_0\;=\left\{{\mathbb E}^4_>\;|\;I_{4,3}^2=I_{4,4}I_{4,1}^2\right\},
&\dim{\mathbb P}^3_0\;=3,\\
{\mathbb P}^4_-=\left\{{\mathbb E}^4_>\;|\;I_{4,3}^2<I_{4,4}I_{4,1}^2\right\},&
\dim{\mathbb P}^4_-=4,\end{array}\right.\label{e26}
\eea
where ${\mathbb P}^3_0$ is a hypersurface, which separates ${\mathbb P}^4_+$ and ${\mathbb P}^4_-$ and is given by equation (\ref{e8}). Formulas (\ref{e23}) and (\ref{e24}) are dual in the following sense:
\bea
\mbox{if}\quad\left(\frac{I_{4,1}I_{4,4}}{I_{4,3}}\right)^{1/2}\le\lambda_
{\odot}\left({\bf x^4}\right)\le\left(\frac{I_{4,3}}{I_{4,1}}\right)^{1/2},\quad
{\bf x^4}\in{\mathbb P}^4_+,\nonumber\\  
\mbox{then}\quad\left(\frac{I_{4,3}}{I_{4,1}}\right)^{1/2}\le\lambda_{\odot}
\left({\bf x^4}\right)\le
\left(\frac{I_{4,1}I_{4,4}}{I_{4,3}}\right)^{1/2},\quad{\bf x^4}\in{\mathbb P}^4_-,\nonumber
\eea
and vice versa, the actual expressions will be exchanged for the lower and upper bounds in subspaces ${\mathbb P}^4_+$ and 
${\mathbb P}^4_-$, while if ${\bf x^4}\in{\mathbb P}^3_0\;$ then 
$\Omega\left({\bf x^4}\right)=\omega\left({\bf x^4}\right)$. 

To compare the upper and lower bounds (\ref{e24}) with the explicit solution (\ref{c10}) in the Mortola-Steff\'e checkerboard, let us find the competitive terms $\sqrt{I_{4,3}/I_{4,1}}$ and 
$\sqrt{I_{4,1}I_{4,4}/I_{4,3}}$ when $x_2=x_4$,
\bea
\sqrt{\frac{I_{4,3}}{I_{4,1}}}=\sqrt{x_2\frac{2x_1x_3+
x_2(x_1+x_3)}{x_1+2x_2+x_3}},\qquad\sqrt{\frac{I_{4,1}I_{4,4}}
{I_{4,3}}}=\sqrt{\frac{x_1x_2x_3\;(x_1+2x_2+x_3)}
{2x_1x_3+x_2(x_1+x_3)}}\;,\label{e27}
\eea
and conclude that $\sqrt{I_{4,3}/I_{4,1}}$ coincides with (\ref{c10}). To determine the asymptotics of the upper and lower bounds in this special setting, we combine (\ref{e24}) and (\ref{e27}),
\bea
&&\hspace{-1cm}
x_1<<x_2,x_3\;:\quad\omega\left({\bf x^4}\right)\simeq
\sqrt{x_1(2x_2+x_3)}\;,\quad\Omega\left({\bf x^4}\right)
\simeq x_2\sqrt{\frac{x_3}{2x_2+x_3}}\;,\nonumber\\
&&\hspace{-1cm}
x_1,x_3<<x_2\;:\quad\omega\left({\bf x^4}\right)\simeq
\sqrt{x_2\;\frac{2x_1 x_3}{x_1+x_3}}\;,\qquad
\Omega\left({\bf x^4}\right)\simeq\sqrt{x_2\frac{x_1+x_3}{2}}\;,\nonumber \\
&&\hspace{-1cm}
x_1,x_2<<x_3\;:\quad\omega\left({\bf x^4}\right)\simeq
\sqrt{x_2(2x_1+x_2)}\;,\quad\;\Omega\left({\bf x^4}\right)\simeq
\sqrt{x_3\;\frac{x_1 x_2}{2x_1+x_2}}\;.\label{e28}
\eea
A brief analysis of (\ref{e28}) shows that in different domains of variables $x_j$ the term $\sqrt{I_{4,3}/I_{4,1}}$ coincides with the upper bound in the first two cases and with the lower bound in the last one.

The formulas (\ref{e28}) give finite, mutually dual bounds in two opposite cases
\bea
\Omega(0,x,x,x)=\frac{x}{\sqrt{3}},\qquad \omega(x,x,x,\infty)=
\sqrt{3}x.\nonumber
\eea
The last formulas predict the effective conductivity of the four-phase Mortola-Steff\'e composite $\;a$) with one insulating and three other similar non-insulating phases and $\;b$) with one superconducting and three other similar non-superconducting phases.

We finish this section by pointing out a compatibility of the upper and lower bounds in (\ref{e24}) with a universal solution $\lambda
\left({\bf x^2}\right)$ of the self-dual equations ${\sf S}\left(^{
\lambda,\;{\bf x^2}}_{m,S_2}\right)=0$, obtained by equating $x_1=
x_3$ and $x_2=x_4$ in (\ref{e24}) and obtaining $\Omega\left(
{\bf x^4}\right)=\omega\left({\bf x^4}\right)=\sqrt{x_1x_2}$.
%%%%%%%%%%%%%%%%%%%%%%%%%%%%%%%%%%%%%%%%%%%%%%%%%%%%%%%%%%%%%%%%%%%%%%%%%%%%%%%%
\section{Comparison of bounds for the effective conductivity}\label{part6}
%%%%%%%%%%%%%%%%%%%%%%%%%%%%%%%%%%%%%%%%%%%%%%%%%%%%%%%%%%%%%%%%%%%%%%%%%%%%%%%%%
In this section, we compare the upper and lower bounds, derived in (\ref{e19}) and (\ref{e24}) for the roots $\lambda_{\odot}\left(
{\bf x^n}\right)$ of the proper self-dual polynomials ${\sf S}_
{\odot}\left(^{\lambda,\;{\bf x^n}}_{m,S_n}\right)$, $n=3,4$, with those, listed in Section \ref{part32} for the effective  conductivity of the 2D three- and four-phase composites, as well as
with the numerical results presented in \cite{fel00}.

$\bullet$
{\sf $\Omega\left({\bf x^n}\right)$ and $\omega\left({\bf x^n}
\right)$ versus solution $\sigma_{Br}\left({\bf x^n}\right)$, $n=
3,4$, of Bruggeman's equation}

\noindent
The Bruggeman equation (\ref{c9}) is a proper, self-dual equation. Based on the correspondence $\mbox{(\ref{e9})}\leftrightarrow
\mbox{(\ref{c9})}$ in (\ref{e10}) and inequalities (\ref{e18}), (\ref{e20}), (\ref{e23}), (\ref{e25}), its solution $\sigma_{Br}\left({\bf x^n}\right)$, $n=3,4$, is bounded
\bea
\omega\left({\bf x^3}\right)\le\sigma_{Br}\left({\bf x^3}\right)\le\Omega\left({\bf x^3}\right),\qquad
\omega\left({\bf x^4}\right)\le\sigma_{Br}\left({\bf x^4}\right)\le\Omega\left({\bf x^4}\right).
\label{f1}
\eea

$\bullet$
{\sf $\Omega\left({\bf x^n}\right)$ and $\omega\left({\bf x^n}
\right)$ versus Wiener bounds $\sigma_{Wi}^{\pm}\left({\bf x^n}\right)$, $n=3,4$}

\noindent
Combining inequalities (\ref{c12}), (\ref{e17}), (\ref{e18}), (\ref{e20}), (\ref{e23}) and (\ref{e25}), we 
obtain
\bea
\sigma_{Wi}^-\left({\bf x^3}\right)\le\omega\left({\bf x^3}\right)\le\sigma_e\left({\bf x^3}\right)\le\Omega\left({\bf x^3}\right)\le
\sigma_{Wi}^+\left({\bf x^3}\right),\label{f2}\\
\sigma_{Wi}^-\left({\bf x^4}\right)\le\omega\left({\bf x^4}\right)
\le\sigma_e\left({\bf x^4}\right)\le\Omega\left({\bf x^4}\right)
\le \sigma_{Wi}^+\left({\bf x^4}\right).\nonumber
\eea

\begin{figure}[h]
\centerline{\psfig{figure=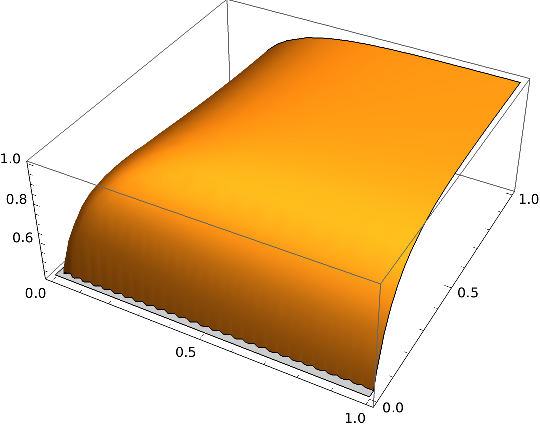,height=5.5cm}}
\caption{Plot of the function $H_3(a,b)$ in the region $0\leq a,b\leq 1$.}\label{fig5}
\end{figure}

$\bullet$
{\sf $\Omega\left({\bf x^3}\right)$ and $\omega\left({\bf x^3}
\right)$ versus Hashin-Shtrickman bounds $\sigma_{HS}^{\pm}\left(
{\bf x^3}\right)$}
\bea
\sigma_{HS}^-\left({\bf x^3}\right)\le \omega\left({\bf x^3}\right)\le\sigma_e\left({\bf x^3}\right)\le\Omega\left({\bf x^3}\right)\le 
\sigma_{HS}^+\left({\bf x^3}\right),\label{f3}
\eea
To prove (\ref{f3}), we represent the formulas for $\sigma_{HS}^{\pm}\left({\bf x^3}\right)$ in (\ref{c15}) for $x_1\le x_2\le x_3$ as follow
\bea
\hspace{-.2cm}
\sigma_{HS}^+\left({\bf x^3}\right)=x_3\;\frac{5 x_1 x_2+3 x_3(x_1+x_2)+x_3^2}{x_1 x_2+3 x_3(x_1+x_2)+5 x_3^2},\quad
\sigma_{HS}^-\left({\bf x^3}\right)=x_1\;\frac{5 x_2 x_3 +3 x_1(x_2+x_3)+ x_1^2}{x_2 x_3+3 x_1(x_2+x_3)+5 x_1^2}.\label{f4}
\eea
A simple algebra gives the differences 
\bea
\sigma_{HS}^+\left({\bf x^3}\right)-\frac{I_{3,2}}{I_{3,1}}&=&
\frac{(x_3-x_1)(x_3 -x_2)(x_3^2-x_1 x_2)}{(x_1+x_2+x_3)
(x_1 x_2+3 x_3(x_1+x_2)+5 x_3^2)}\geq 0\;,\label{f5}\\
\frac{I_{3,2}}{I_{3,1}}-\sigma_{HS}^-\left({\bf x^3}\right)&=&
\frac{(x_2 - x_1) (x_3 - x_1) (x_2 x_3 - x_1^2)}{(x_1+x_2+x_3)
(x_2 x_3+3 x_1(x_2+x_3)+5 x_1^2)}\geq 0\;,\nonumber
\eea
and the ratios
\bea
\sigma_{HS}^-\left({\bf x^3}\right)\left(\frac{I_{3,3}I_{3,1}}{I_{3,2}}\right)^{-1/2}=H_3\left(\frac{x_1}{x_2},\frac{x_1}{x_3}\right),\qquad
\left(\frac{I_{3,3}I_{3,1}}{I_{3,2}}\right)^{1/2}
\frac1{\sigma_{HS}^+\left({\bf x^3}\right)}=H_3\left(
\frac{x_1}{x_3},\frac{x_2}{x_3}\right),\label{f6}
\eea
where
\bea
H_3(a,b)=\left(ab\;\frac{1+a+b}{a+b+ab}\right)^{1/2}\frac{
5+3(a+b)+ab}{1+3(a+b)+5ab},\qquad 0\le a,b\le 1.\label{f7}
%ab\left(\frac{1+a+b}{1+a^{-1}+b^{-1}}\right)
%^{1/2}\frac{1+3(a^{-1}+b^{-1})+5a^{-1}b^{-1}}{1+3(a+b)+5ab}
\eea
In Figure \ref{fig5}, we present a plot of the function $H_3(a,b)$ that shows that $H_3(a,b)\le 1$ when $0\le a,b\le 1$. Combining this inequality with (\ref{f5}, \ref{f6}), we get
\bea
\sigma_{HS}^-\left({\bf x^3}\right)\le\frac{I_{3,2}}{I_{3,1}},
\left(\frac{I_{3,3}I_{3,1}}{I_{3,2}}\right)^{1/2}\quad\mbox{and}\qquad
\sigma_{HS}^+\left({\bf x^3}\right)\ge\frac{I_{3,2}}{I_{3,1}},
\left(\frac{I_{3,3}I_{3,1}}{I_{3,2}}\right)^{1/2},\label{f8}
\eea
which proves (\ref{f3}). In Figure \ref{fig7} we present different cross-sections of the 2D plot of the universal bounds $\omega
\left({\bf x^3}\right)$ and $\Omega\left({\bf x^3}\right)$ in comparison with the Wiener, Hashin-Shtrickman and Nesi bounds.

$\bullet$
{\sf $\Omega\left({\bf x^3}\right)$ and $\omega\left({\bf x^3}
\right)$ versus Nesi's bounds $\sigma_{Ne}^{\pm}\left({\bf x^3}\right)$}
\bea
\sigma_{Ne}^-\left({\bf x^3}\right)\le\omega\left({\bf x^3}\right)\le\sigma_e\left({\bf x^3}\right)\le\Omega\left({\bf x^3}\right)\le
\sigma_{Ne}^+\left({\bf x^3}\right).\label{f9}
\eea
To prove inequalities (\ref{f9}), represent formulas for the real solutions $\sigma_{Ne}^{\pm}\left({\bf x^3}\right)$ of cubic equations (\ref{c18}) 
\bea
&&\hspace{-.5cm}
\sigma_{Ne}^-\left({\bf x^3}\right)=\frac{I_{3,1}}{3}\left(\frac{
I_{3,1}}{\sqrt[3]{U_{Ne}^-}}+\frac{\sqrt[3]{U_{Ne}^-}}{I_{3,1}}-1
\right),\quad U_{Ne}^-=-I_{3,1}^3+54 I_{3,3}+6\sqrt{3}
\sqrt{27 I_{3,3}^2-I_{3,1}^3 I_{3,3}},\nonumber\\
&&\hspace{-.5cm}
\sigma_{Ne}^+\left({\bf x^3}\right)=\frac{1}{2\sqrt[3]{3}}\left(
\frac{I_{3,2}}{\sqrt[3]{U_{Ne}^+}}+\sqrt[3]{\frac{U_{Ne}^+}{3}}\;\right),\quad U_{Ne}^+=9I_{3,3}+\sqrt{3}\sqrt{27 I_{3,3}^2-
I_{3,2}^3},\quad x_1\le x_2\le x_3.\nonumber
\eea
Consider the ratios
\bea
&&\sigma_{Ne}^-\left({\bf x^3}\right)\left(\frac{I_{3,2}}{I_{3,1}}\right)^{-1}=N_3^{1-}\left(x_2,x_3\right),\qquad
\sigma_{Ne}^-\left({\bf x^3}\right)\left(\frac{I_{3,3}I_{3,1}}{I_{3,2}}\right)^{-1/2}=N_3^{2-}\left(x_2,x_3\right),\nonumber\\
&&\frac{I_{3,2}}{I_{3,1}}\frac1{\sigma_{Ne}^+\left({\bf x^3}\right)}=N_3^{1+}\left(x_2,x_3\right),\qquad\left(\frac{I_{3,3}
I_{3,1}}{I_{3,2}}\right)^{1/2}\frac1{\sigma_{Ne}^+\left({\bf x^3}
\right)}=N_3^{2+}\left(x_2,x_3\right),\label{f10}
\eea

\begin{figure}[h!]\begin{center}\begin{tabular}{ccc}
\psfig{figure=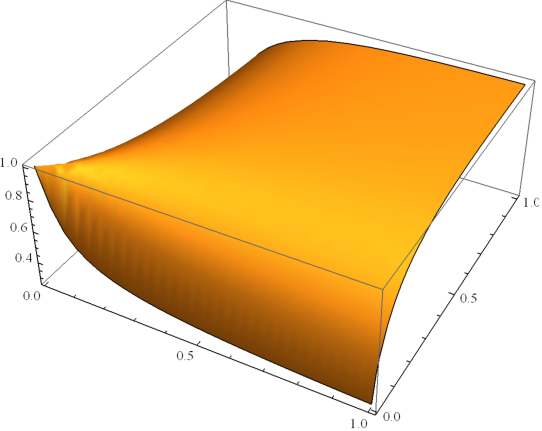,height=4.7cm}
\hspace{.2cm}&\hspace{.2cm}&\hspace{.2cm}
\psfig{figure=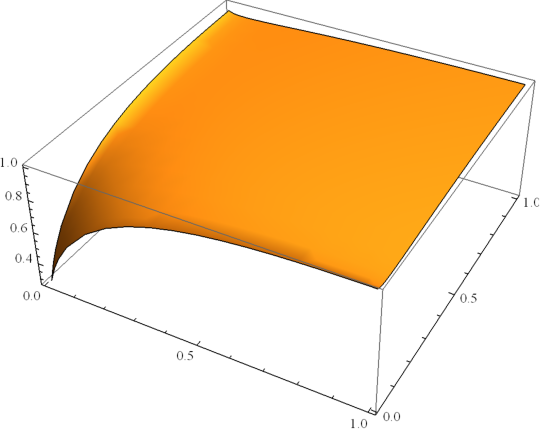,height=4.7cm}
\\ (1) & & (2)\\
\psfig{figure=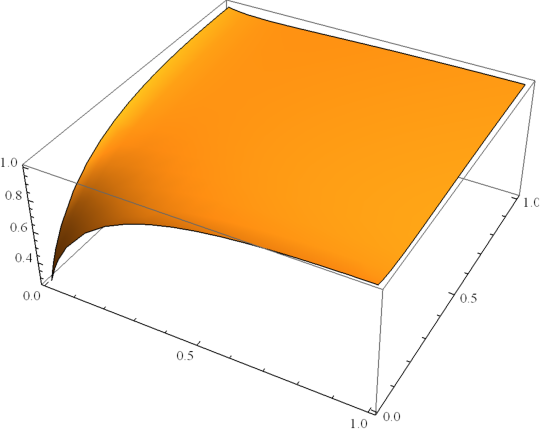,height=4.7cm}
\hspace{.2cm}&\hspace{.2cm}&\hspace{.2cm}
\psfig{figure=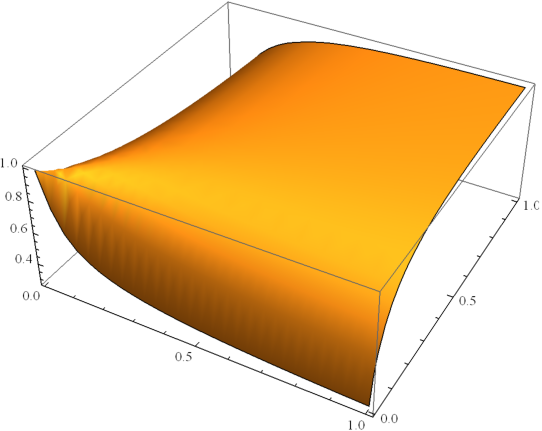,height=4.7cm}
\\ (3) & & (4)
\end{tabular}\end{center}
\vspace{-.3cm}
\caption{Typical plot of the functions 1) $N_3^{1-}(a,b),\;$ 2) $N_3^{2-}(a,b),\;$ 3) $N_3^{1+}(a,b)$ and 4) $N_3^{2+}(a,b)$ in the region $0\le\!a,b\!\le 1$.}\label{fig6}
\end{figure}

\begin{figure}[h!]\begin{center}\begin{tabular}{ccc}
\psfig{figure=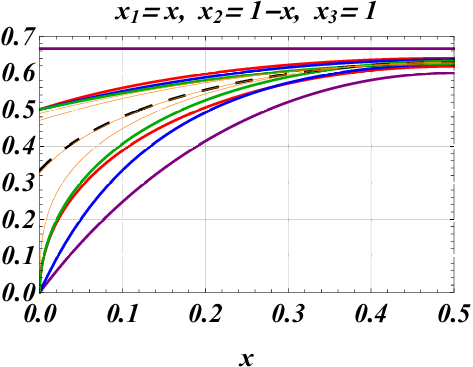,height=5.cm}
\hspace{-.1cm}&\hspace{-.1cm}&\hspace{-.1cm}
\psfig{figure=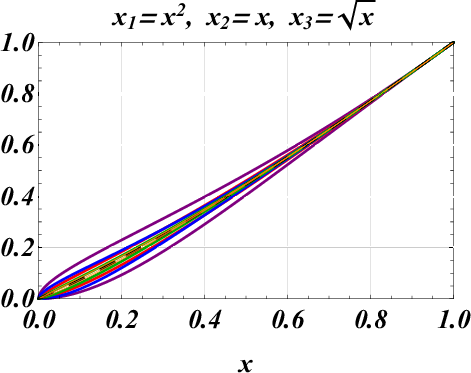,height=5.cm}
\\ 
\psfig{figure=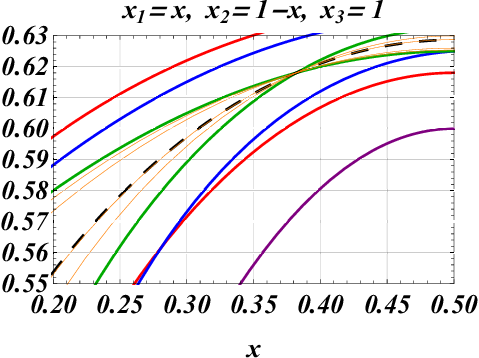,height=5.cm}
\hspace{-.1cm}&\hspace{-.1cm}&\hspace{-.1cm}
\psfig{figure=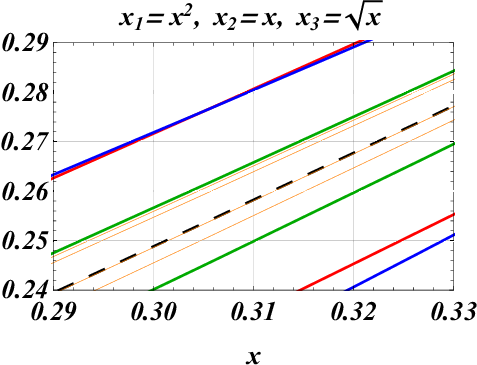,height=5.cm}
\\ 
\psfig{figure=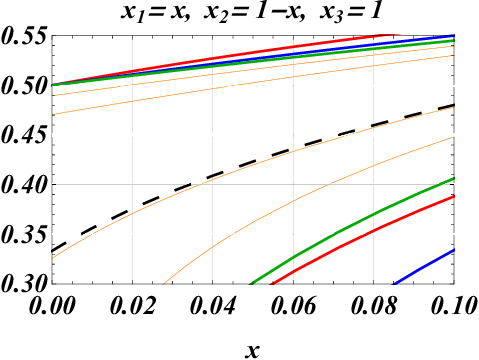,height=5.cm}
\hspace{-.1cm}&\hspace{-.1cm}&\hspace{-.1cm}
\psfig{figure=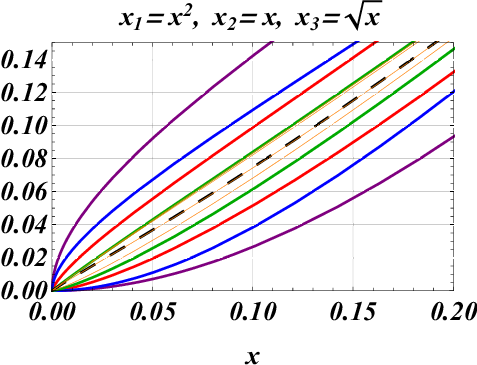,height=5.cm}
\end{tabular}\end{center}
\vspace{-.3cm}
\caption{Two cross-sections of the universal surfaces corresponding to the upper and lower bounds: Wiener's bounds $\sigma_{Wi}^{\pm}\left({\bf x^3}\right)$ -{\it brown}, Nesi's bounds $\sigma_{Ne}^
{\pm}\left({\bf x^3}\right)$ -{\it red}, Hashin-Shtrikman's bounds $\sigma_{HS}^{\pm}\left({\bf x^3}\right)$ - {\it blue}, upper and lower bounds $\Omega\left({\bf x^3}\right)$ and $\omega\left(
{\bf x^3}\right)$ - {\it green}. Bruggeman's solution $\sigma_{Br}
\left({\bf x^3}\right)$ - {\it dashed black}, the structures {\sf He}, {\sf Fl}, {\sf Co} and {\sf Rh} - {\it orange}. Detailed fragments show the most critical regions of bounds coexistence (Dykhne' ansatz): Bruggeman's solution passes through the intersection of the upper and lower bounds (Dykhne's points). Note that Hashin-Shtrikman's and Nesi's bounds are intersecting.
}\label{fig7}
\end{figure}

Rather than using a sophisticated algebraic approach, we present in Figure \ref{fig6} the numerical calculations of functions $N_3^{k
\pm}\left(x_2,x_3\right)$, $k=1,2$. These functions satisfy inequalities $N_3^{k\pm}\left(x_2,x_3\right)\le 1$ provided that $x_2,x_3\le 1$. By combining these inequalities with expressions (\ref{f5} and \ref{f6}), we get
\bea
\sigma_{Ne}^-\left({\bf x^3}\right)\le\frac{I_{3,2}}{I_{3,1}},
\left(\frac{I_{3,3}I_{3,1}}{I_{3,2}}\right)^{1/2}\quad\mbox{and}\qquad
\sigma_{Ne}^+\left({\bf x^3}\right)\ge\frac{I_{3,2}}{I_{3,1}},
\left(\frac{I_{3,3}I_{3,1}}{I_{3,2}}\right)^{1/2},\label{f11}
\eea
that proves (\ref{f9}).
%%%%%%%%%%%%%%%%%%%%%%%%%%%%%%%%%%%%%%%%%%%%%%%%%%%%%%%%%%
%%%%%%%%%%%%%%%%%%%%%%%%%%%%%%%%%%%%%%%%%%%%%%%%%%%%%%%%%%

$\bullet$
{\sf $\Omega\left({\bf x^3}\right)$ and $\omega\left({\bf x^3}
\right)$ and numerical results for three-phase {\sf He}, {\sf Fl}, {\sf Co}, {\sf Rh} composites \cite{fel00}}

\noindent
In Figure \ref{fig7}, all four plots for the different structures 
{\sf He}, {\sf Fl}, {\sf Co} and {\sf Rh}, numerically calculated in \cite{fel00}, are located between the upper and lower bounds 
$\Omega\left({\bf x^3}\right)$ and $\omega\left({\bf x^3}\right)$ and pass through their intersection.
%%%%%%%%%%%%%%%%%%%%%%%%%%%%%%%%%%%%%%%%%%%%%%%%%%%%%%%%%%

$\bullet$
{\sf $\Omega\left({\bf x^3}\right)$ and $\omega\left({\bf x^3}
\right)$ versus Cherkaev's bounds $\sigma_{Ch}^{\pm}\left({\bf x^3}
\right)$, $\;x_1\le x_2\ll x_3$}
\bea
\sigma_{Ch}^-\left({\bf x^3}\right)\le\omega\left({\bf x^3}\right)\le\sigma_e\left({\bf x^3}\right)\le\Omega\left({\bf x^3}\right)\le 
\sigma_{Ch}^+\left({\bf x^3}\right),\label{f12}
\eea

To prove (\ref{f12}), we consider a three-phase compound with one almost superconducting constituent $x_3\gg x_2\ge x_1$. According to (\ref{e19}), we get
\bea
\omega(x_1,x_2,x_3)\simeq x_1+x_2,\qquad\Omega(x_1,x_2,x_3)\simeq \sqrt{\frac{x_1x_2}{x_1+x_2}}\sqrt{x_3},\label{f13}
\eea
Comparing (\ref{f13}) with $\sigma_{Ch}^-\left({\bf x^3}\right)$ 
in (\ref{c24}) and $\sigma_{Ch}^+\left({\bf x^3}\right)$ in (\ref{c26}), we prove (\ref{f12}) :
\bea
\omega(x_1,x_2,x_3)\!-\!\sigma_{Ch}^-(x_1,x_2,x_3)=\frac{x_2(x_2-x_1)}{{3x_1+x_2}}\ge 0,\quad\frac{\sigma_{Ch}^+(x_1,x_2,x_3)}{\Omega(x_1,x_2,x_3)}\simeq \frac1{5}\sqrt{\frac{x_1+x_2}{x_2}}\sqrt{\frac{x_3}{x_1}}\gg 1.\nonumber
\eea
%%%%%%%%%%%%%%%%%%%%%%%%%%%%%%%%%%%%%%%%
%%%%%%%%%%%%%%%%%%%%%%%%%%%%%%%%%%%%%%%%%
\subsection{$\Omega\left({\bf x^4}\right)$ and $\omega\left(
{\bf x^4}\right)$ versus Hashin-Shtrickman's bounds $\sigma_{HS}^
{\pm}\left({\bf x^4}\right)$}\label{part61}
%%%%%%%%%%%%%%%%%%%%%%%%%%%%%%%%%%%%%%%%%%%%%%%%
%%%%%%%%%%%%%%%%%%%%%%%%%%%%%%%%%%%%%%%%%%%%%%%%
\bea
\sigma_{HS}^-\left({\bf x^4}\right)\le \omega\left({\bf x^4}\right)\le\sigma_e\left({\bf x^4}\right)\le\Omega\left({\bf x^4}\right)\le 
\sigma_{HS}^+\left({\bf x^4}\right),
\label{f14}
\eea
To prove (\ref{f14}), we present the expressions for $\sigma_{HS}^
{\pm}\left({\bf x^4}\right)$ in the region $x_1 \geq x_2 \geq x_3
\geq x_4$ in accordance with the formulas (\ref{c15}),
\bea
\sigma_{HS}^+\left({\bf x^4}\right)=x_1\;\frac{7 x_2 x_3 x_4+5 x_1
(x_2 x_3+x_3 x_4+x_4 x_2)+3 x_1^2(x_2+x_3+x_4)+x_1^3}{x_2x_3x_4+
3x_1(x_2 x_3+x_3 x_4+x_4x_2)+5x_1^2(x_2+x_3+x_4)+7x_1^3}\;,
\label{f15}\\
\sigma_{HS}^-\left({\bf x^4}\right)=x_4\;\frac{7x_1x_2x_3+5x_4
(x_1x_2+x_2x_3+x_3x_1)+3x_4^2(x_1+x_2+x_3)+x_4^3}{x_1x_2x_3+3x_4
(x_1x_2+x_2x_3+x_3x_1)+5x_4^2(x_1+x_2+x_3)+7x_4^3}\;.\nonumber
\eea
%%%%%%%%%%%%%%%%%%%%%%%%%%%%%%%%%%%%%%%%%%%%%%%%%%%%%%%%%%
\begin{figure}[h!]\begin{center}\begin{tabular}{ccc}
\psfig{figure=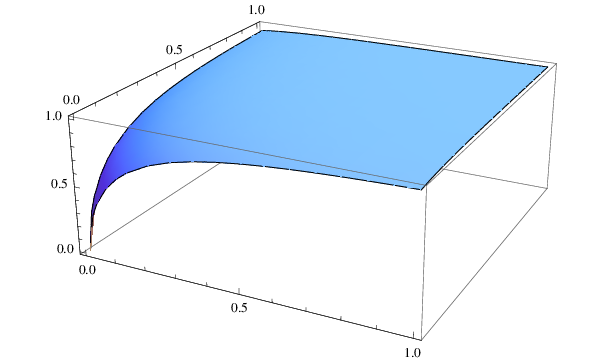,height=4.5cm}
\hspace{.05cm}&\hspace{.05cm}&\hspace{.05cm}
\psfig{figure=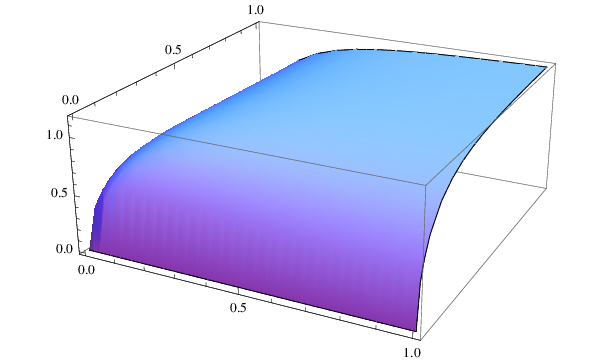,height=4.5cm}
%\\ (1) & & (2)
\end{tabular}\end{center}
\vspace{-.3cm}
\caption{Typical plot of functions $H_{4,1}(a,b,1)$ ($left$) and $H_{4,2}(a,b,1)$ ($right$) in the region $0\le\!a,b\!\le 1$.}\label{fig8}
\end{figure}

Define the ratios
\bea
\sigma_{HS}^-\left({\bf x^4}\right)\left(\frac{I_{4,3}}{I_{4,1}}\right)^{-1/2}\!\!\!\!=H_{4,1}\left(\frac{x_4}{x_1},\frac{x_4}{x_2},\frac{x_4}{x_3}\right),\qquad\sigma_{HS}^-\left({\bf x^4}
\right)\left(\frac{I_{4,4}I_{4,1}}{I_{4,3}}\right)^{-1/2}\!\!\!\!=H_{4,2}\left(\frac{x_4}{x_1},\frac{x_4}{x_2},\frac{x_4}{x_3}\right),\nonumber\\
\left(\frac{I_{4,3}}{I_{4,1}}\right)^{1/2}\!\!\frac1{\sigma_{HS}^+
\left({\bf x^4}\right)}=H_{4,1}\left(\frac{x_2}{x_1},\frac{x_3}
{x_1},\frac{x_4}{x_1}\right),\qquad\left(\frac{I_{4,4}I_{4,1}}{I_{4,3}}\right)^{1/2}\!\!\frac1{\sigma_{HS}^+\left({\bf x^4}
\right)}=H_{4,2}\left(\frac{x_2}{x_1},\frac{x_3}{x_1},\frac{x_4}
{x_1}\right),\nonumber
\eea
where
\bea
&&
H_{4,1}(a_1,a_2,a_3)=\left(\frac{a_1a_2+a_2a_3+a_3a_1+a_1a_2a_3}{1+a_1+a_2+a_3}\right)^{1/2}V_4(a_1,a_2,a_3),\label{f16}\\
&&
H_{4,2}(a_1,a_2,a_3)=\left(\frac{a_1a_2a_3(1+a_1+a_2+a_3)}{a_1a_2+a_2a_3+a_3a_1+a_1a_2a_3}\right)^{1/2}
V_4(a_1,a_2,a_3),\nonumber\\
&&
V_4(a_1,a_2,a_3)=\frac{7+5(a_1+a_2+a_3)+3(a_1a_2+a_2a_3+a_3a_1)+a_1a_2a_3}{1+3(a_1+a_2+a_3)+5(a_1a_2+a_2a_3+a_3a_1)+7a_1a_2a_3}\;,\nonumber
\eea

\begin{figure}[h!]\begin{center}\begin{tabular}{ccc}
\psfig{figure=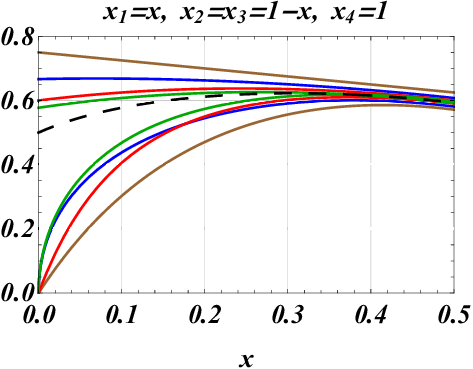,height=5.cm}
\hspace{-.1cm}&\hspace{-.1cm}&\hspace{-.1cm}
\psfig{figure=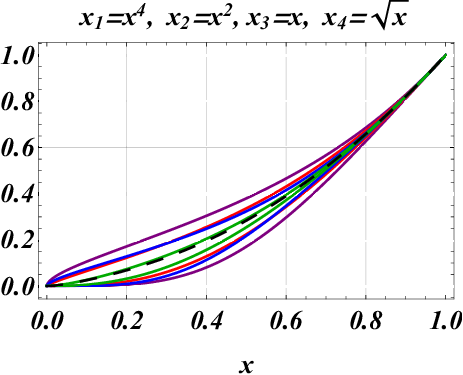,height=5.cm}
\\ 
\psfig{figure=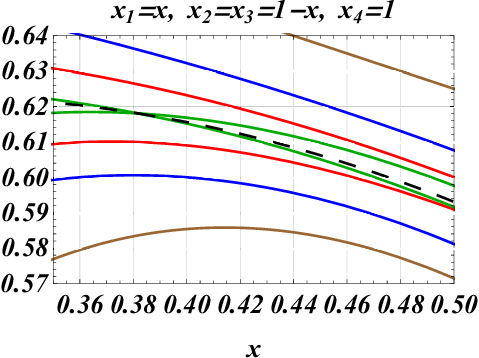,height=5.cm}
\hspace{-.1cm}&\hspace{-.1cm}&\hspace{-.1cm}
\psfig{figure=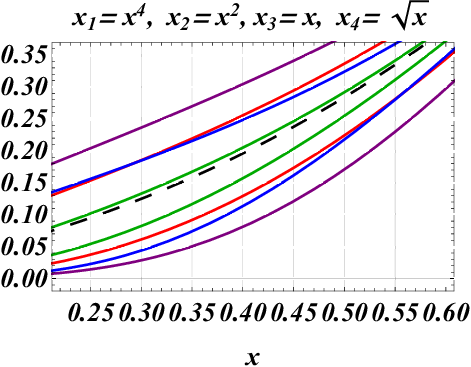,height=5.cm}
\\ 
\psfig{figure=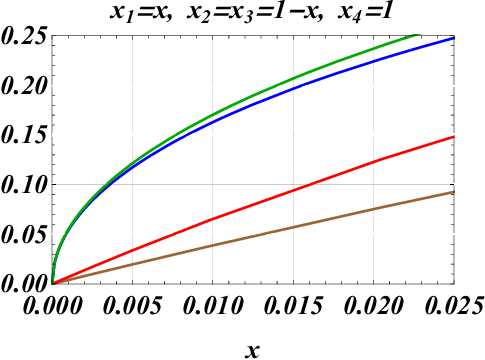,height=5.cm}
\hspace{-.1cm}&\hspace{-.1cm}&\hspace{-.1cm}
\psfig{figure=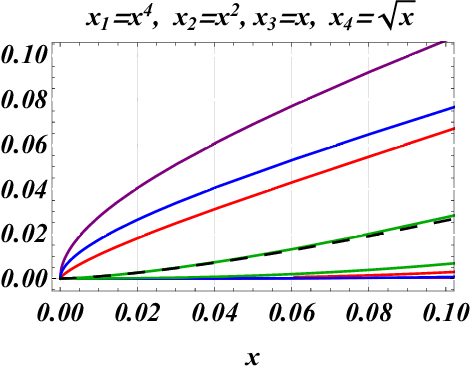,height=5.cm}
\end{tabular}\end{center}
\vspace{-.3cm}
\caption{Two cross-sections of the universal surfaces corresponding to the upper and lower bounds: Wiener's bounds $\sigma_
{Wi}^{\pm}\left({\bf x^4}\right)$ - {\it brown}, conjectured bounds $\sigma_{Cj}^{\pm}\left({\bf x^4}\right)$ - {\it red}, Hashin-Shtrikman's bounds $\sigma_{HS}^{\pm}\left({\bf x^4}\right)$  - {\it blue}, upper and lower bounds $\Omega\left({\bf x^4}\right)$ and $\omega\left({\bf x^4}\right)$ - {\it green}. Bruggeman's solution $\sigma_{Br}\left({\bf x^4}\right)$ - {\it dashed black}. Detailed fragments show the most critical regions of the bounds coexistence (Dykhne's ansatz): Bruggeman's solution passes through the intersection of the upper and lower bounds.}\label{fig9}
\end{figure}

It is easy to check numerically that both functions $H_{4,j}(a,b,c)$ satisfy $\partial H_{4,j}/\partial a_k>0$, $k=1,2,3$, which results in $0\le H_{4,j}(a_1,a_2,a_3)\le 1$ (see Figure \ref{fig8}). Keeping in mind the definition (\ref{f16}), we get
\bea
\sigma_{HS}^-\left({\bf x^4}\right)\le\left(\frac{I_{4,3}}{I_{4,1}}\right)^{1/2}\!\!,\left(\frac{I_{4,4}I_{4,1}}{I_{4,3}}\right)^{1/2}
\quad\mbox{and}\qquad\sigma_{HS}^+\left({\bf x^4}\right)\ge
\left(\frac{I_{4,3}}{I_{4,1}}\right)^{1/2}\!\!,\left(\frac{I_{4,4}I_{4,1}}{I_{4,3}}\right)^{1/2},\label{f17}
\eea
that proves (\ref{f14}). Figure \ref{fig9} presents two 
cross-sections of the universal bounds.

%%%%%%%%%%%%%%%%%%%%%%%%%%%%%%%%%%%%%%%%%%%%%%%%%%%%%%%%%%
%%%%%%%%%%%%%%%%%%%%%%%%%%%%%%%%%%%%%%%%%%%
%%%%%%%%%%%%%%%%%%%%%%%%%%%%%%%%%%%%%%%%%
\subsection{$\Omega\left({\bf x^4}\right)$ and $\omega\left(
{\bf x^4}\right)$ versus conjectured bounds $\sigma_{Cj}^{\pm}
\left({\bf x^4}\right)$}\label{part62}
%%%%%%%%%%%%%%%%%%%%%%%%%%%%%%%%%%%%%%%%%%%%%%%%
%%%%%%%%%%%%%%%%%%%%%%%%%%%%%%%%%%%%%%%%%%%%%%%%
\bea
\sigma_{Cj}^-\left({\bf x^4}\right)\le \omega\left({\bf x^4}\right)\le\sigma_e\left({\bf x^4}\right)\le\Omega\left({\bf x^4}\right)\le \sigma_{Cj}^+\left({\bf x^4}\right),\label{f18}
\eea
To prove inequalities (\ref{f18}), we present formulas for real solutions $\sigma_{Cj}^{\pm}\left({\bf x^3}\right)$ of quartic equations (\ref{c19})
\bea
&&\hspace{-.5cm}
2\sigma_{Cj}^-\left({\bf x^4}\right)=-I_{4,1}+\sqrt{I_{4,1}^2-2
I_{4,2}+A_+}+\sqrt{2(I_{4,1}^2-2I_{4,2})-A_+-\frac{2I_{4,1}(I_{4,1}
^2-3I_{4,2})}{\sqrt{I_{4,1}^2-2I_{4,2}+A_+}}}\;,\nonumber\\
&&\hspace{-.5cm}
6\sqrt{2}\sigma_{Cj}^+\left({\bf x^4}\right)=\sqrt{2I_{4,2}+A_-}+
\sqrt{4I_{4,2}-A_-+\frac{12\sqrt{3}\;I_{4,3}}{\sqrt{2I_{4,2}+A_-}}}\;,\quad\mbox{where}\nonumber\\
&&\hspace{-.5cm}
A_{\pm}=\beta\;\alpha_{\pm}^{-1/3}+\alpha_{\pm}^{1/3},\;\;
\alpha_{\pm}=\sqrt{\gamma^2-\beta^3}\pm\gamma,\;\;
\beta=I_{4,2}^2-36I_{4,4},\;\;\gamma=I_{4,2}^3-I_{4,1}^2I_{4,4}+I_{4,2}I_{4,4}.\nonumber
\eea
%%%%%%%%%%%%%%%%%%%%%%%%%%%%%%%%%%%%%%%%%%%%%%%%%%%%%%%%%%
\begin{figure}[h!]\begin{center}\begin{tabular}{ccc}
\psfig{figure=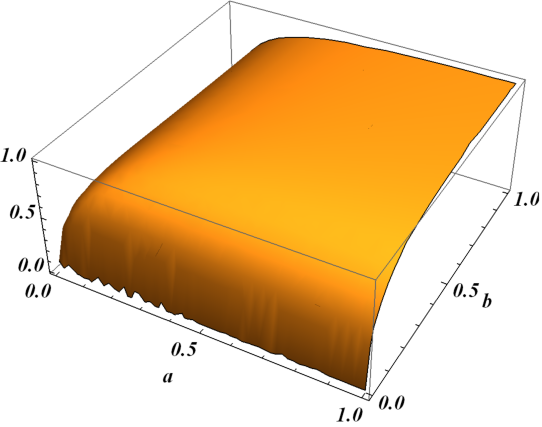,height=4.6cm}
\hspace{.2cm}&\hspace{.2cm}&\hspace{.2cm}
\psfig{figure=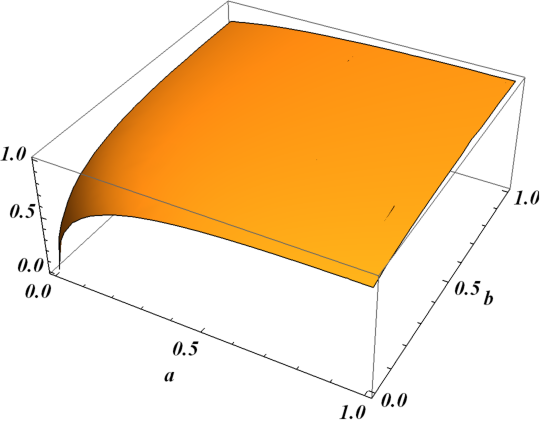,height=4.6cm}
\\ (1) & & (2)\\
\psfig{figure=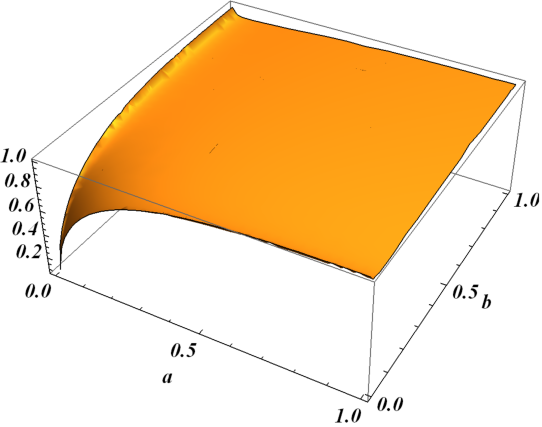,height=4.6cm}
\hspace{.2cm}&\hspace{.2cm}&\hspace{.2cm}
\psfig{figure=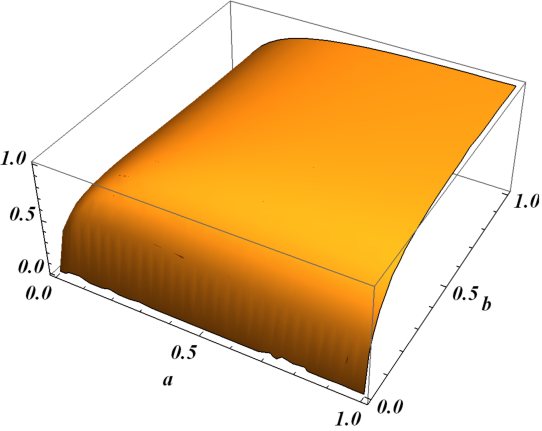,height=4.6cm}
\\ (3) & & (4)
\end{tabular}\end{center}
\vspace{-.3cm}
\caption{Typical plot of the functions 1) $U_{4,1}^-(a,b,1)$, 2) $U_{4,1}^+(a,b,1)$, 3) $U_{4,2}^-(a,b,1)$ and 4) $U_{4,2}^+(a,b,1)$ in the region $0\le\!a,b\!\le 1$.}\label{fig10}
\end{figure}
%%%%%%%%%%%%%%%%%%%%%%%%%%%%%%%%%%%%%%%%%%%%%%%%%%%%%%%%%%

Assuming that $x_1\le x_2\le x_3\le x_4$, define the ratios
\bea
&&\hspace{-1.5cm}
\sigma_{Cj}^-\left({\bf x^4}\right)\left(\frac{I_{4,3}}{I_{4,1}}\right)^{-1/2}\!\!\!\!=U_{4,1}^-\left(\frac{x_4}{x_1},\frac{x_4}{x_2},\frac{x_4}{x_3}\right),\quad\sigma_{Cj}^-\left({\bf x^4}
\right)\left(\frac{I_{4,4}I_{4,1}}{I_{4,3}}\right)^{-1/2}\!\!\!\!=U_{4,2}^-\left(\frac{x_4}{x_1},\frac{x_4}{x_2},\frac{x_4}{x_3}\right),\nonumber\\
&&\hspace{-1.5cm}
\left(\frac{I_{4,3}}{I_{4,1}}\right)^{1/2}\!\!\frac1{\sigma_{Cj}^+
\left({\bf x^4}\right)}=U_{4,1}^+\left(\frac{x_2}{x_1},\frac{x_3}
{x_1},\frac{x_4}{x_1}\right),\quad\left(\frac{I_{4,4}I_{4,1}}{I_{4,3}}\right)^{1/2}\!\!\frac1{\sigma_{Cj}^+\left({\bf x^4}
\right)}=U_{4,2}^+\left(\frac{x_2}{x_1},\frac{x_3}{x_1},\frac{x_4}
{x_1}\right).\quad\label{f19}
\eea
Rather than a sophisticated algebra, we present in Figures \ref{fig10} the numerical calculations of functions $U_{4,j}^{\pm}
\left(a,b,1\right)$, $j\!=\!1,2$, which satisfy $U_{4,j}^{\pm}\left(a,b,1\right)\le 1$ if $a,b\le 1$. Combining this with (\ref{f19}), we get
\bea
\sigma_{Cj}^-\left({\bf x^4}\right)\le\left(\frac{I_{4,3}}{I_{4,1}}\right)^{1/2}\!\!,\left(\frac{I_{4,4}I_{4,1}}{I_{4,3}}\right)^{1/2}
\quad\mbox{and}\qquad\sigma_{Cj}^+\left({\bf x^4}\right)\ge
\left(\frac{I_{4,3}}{I_{4,1}}\right)^{1/2}\!\!,\left(\frac{I_{4,4}I_{4,1}}{I_{4,3}}\right)^{1/2},\label{f20}
\eea
that proves (\ref{f18}).
%%%%%%%%%%%%%%%%%%%%%%%%%%%%%%%%%%%%%%%%%%%%%%%
%%%%%%%%%%%%%%%%%%%%%%%%%%%%%%%%%%%%%%%%%%%
\section{Open problems}\label{part7}
%%%%%%%%%%%%%%%%%%%%%%%%%%%%%%%%%%%%%%%%%%%%%%%
%%%%%%%%%%%%%%%%%%%%%%%%%%%%%%%%%%%%%%%%%%%
We put forward two mathematical conjectures that essentially generalize the statements in Sections \ref{part3} and \ref{part4}.
The first conjecture extends the determinant equations (\ref{c17}) for the bounds $\sigma_{Cj}^{\pm}(\sigma_1,\ldots,\sigma_n)$ to an arbitrary $n\ge 4$.
%%%%%%%%%%%%%%%%%%%%%%%%%%%%%%%%%%%%%%%%%%%
\begin{conjecture}\label{con1}
{\em The following equations provide the lower $\sigma_{Cj}^-=
\sigma_{Cj}^-(\sigma_1,\ldots,\sigma_n)$ and upper $\sigma_{Cj}^+=\sigma_{Cj}^+(\sigma_1,\ldots,\sigma_n)$ bounds for effective conductivity of the 2D $n$-phase composite
\bea
&&
\det\left(\begin{array}{rrrr}
(n-1)\sigma_1 & -\sigma_{Cj}^- & \ldots & -\sigma_{Cj}^-\\
-\sigma_{Cj}^- & (n-1)\sigma_2 & \ldots & -\sigma_{Cj}^-\\
\ldots & \ldots& \ldots & \ldots\\
-\sigma_{Cj}^- & -\sigma_{Cj}^-&\ldots &(n-1)\sigma_n\end{array}\right)=0,\label{g1}\\\nonumber\\
&&
\det\left(\begin{array}{rrrr}
(n-1)/\sigma_1& -1/\sigma_{Cj}^+& \ldots & -1/\sigma_{Cj}^+\\
-1/\sigma_{Cj}^+ & (n-1)/\sigma_2& \ldots & -1/\sigma_{Cj}^+\\
\ldots & \ldots &\ldots & \ldots \\
-1/\sigma_{Cj}^+&-1/\sigma_{Cj}^+&\ldots &(n-1)/\sigma_n\end{array}\right)=0.\label{g2}
\eea}
\end{conjecture}
%%%%%%%%%%%%%%%%%%%%%%%%%%%%%%%%%%%%%%%%%%%
Two determinant equations may be represented as polynomial equations
\bea
&&\hspace{-1cm}
\left(\sigma_{Cj}^-\right)^n+\sum_{k=0}^{n-2}(n-k-2)(n-1)^k
I_{n,k+1}\left(\sigma_{Cj}^-\right)^{n-k-1}-(n-1)^{n-1}I_{n,n}=0,
\label{g3}\\
&&\hspace{-1cm}
\left(\sigma_{Cj}^+\right)^n-\sum_{k=n-2}^{0}\frac{n-k-2}{(n-1)^
{n-k-1}}I_{n,n-k-1}\left(\sigma_{Cj}^+\right)^{k+1}-I_{n,n}
\frac1{(n-1)^{n-1}}=0.\label{g4}
\eea
The functions $\sigma_{Cj}^{\pm}(\sigma_1,\ldots,\sigma_n)$ are permutation-invariant and mutually dual.

Next, we pose the conjecture generalizing the existence of a sequence (\ref{d23}) of embedded spaces for self-dual polynomials
${\sf S}\left(^{\lambda,\;{\bf x^3}}_{m,S_3}\right)$ and ${\sf S}\left(^{\lambda,\;{\bf x^4}}_{m,S_4}\right)$ to polynomials
${\sf R}\left(^{\lambda,\;{\bf x^n}}_{m,S_n}\right)$ and ${\sf S}\left(^{\lambda,\;{\bf x^n}}_{m,S_n}\right)$ with arbitrary $n$.
%%%%%%%%%%%%%%%%%%%%%%%%%%%%%%%%%%%%%%%%%%%
\begin{conjecture}\label{con21}
{\em Let ${\cal M}_{RR}^R\left(^{\lambda,\;{\bf x^n}}_{m,S_n}\right)$, ${\cal M}_{SS}^R\left(^{\lambda,\;{\bf x^n}}_{m,S_n}
\right)$ and ${\cal M}_{RS}^S\left(^{\lambda,\;{\bf x^n}}_{m,S_n}
\right)$ be the differences of unimodality numbers, as defined in (\ref{d15}). Then, for $m_1,m_2\ge 1$ and $n\ge 3$ the following holds:}
\bea
{\cal M}_{RR}^R\left(^{\lambda,\;{\bf x^n}}_{m,S_n}\right)>0,\qquad
{\cal M}_{SS}^R\left(^{\lambda,\;{\bf x^n}}_{m,S_n}\right)>0,\qquad
{\cal M}_{RS}^S\left(^{\lambda,\;{\bf x^n}}_{m,S_n}\right)>0.
\label{g5}
\eea
\end{conjecture}
%%%%%%%%%%%%%%%%%%%%%%%%%%%%%%%%%%%%%%%%%%%%%%%
%%%%%%%%%%%%%%%%%%%%%%%%%%%%%%%%%%%%%%%%%%%%%%%
\section{Concluding remarks}\label{part8}
%%%%%%%%%%%%%%%%%%%%%%%%%%%%%%%%%%%%%%%%%%%%%%%
%%%%%%%%%%%%%%%%%%%%%%%%%%%%%%%%%%%%%%%%%%%
We applied an algebraic approach, developed within the framework of the theory of commutative monoid of self-dual polynomials \cite{fel22}, to the problem of the isotropic effective conductivity $\sigma_e(\sigma_1,\ldots,\sigma_n)$ in the 2D three- and four-phase symmetric composites with partial isotropic conductivities $\sigma_j$.

We derived the upper $\Omega_n=\Omega(\sigma_1,\ldots,\sigma_n)$ and lower $\omega_n=\omega(\sigma_1,\ldots,\sigma_n)$, $n=3,4$, bounds, which are universal, that is, they are independent of the composite microstructure. They possess all the algebraic properties of $\sigma_e(\sigma_1,\ldots,\sigma_n)$ that follow from physics: first-order homogeneity, full permutation invariance, self-duality, and monotony. This is a consequence of two properties,  self-duality and $S_n$-permutation invariance of self-dual polynomials, which are both inherited by $\Omega_n$- and $\omega_n$- bounds. This stands in contrast to the Hashin-Shtrikman bounds $\sigma_{HS}^{\pm}(\sigma_1,\ldots,\sigma_n)$ in (\ref{c15}, \ref{f4}, \ref{f15}) (which are self-dual, but not fully permutation invariant) and Nesi's (and  conjectured) bounds $\sigma_{N,Cj}^
{\pm}(\sigma_1,\ldots,\sigma_n)$ in (\ref{c18}, \ref{c19},\ref{g3}, \ref{g4}) (which are fully permutation invariant, but not self-dual).

The bounds $\Omega_n$ and $\omega_n$ are compatible with the trivial solution $\sigma_e(\sigma,\ldots,\sigma)=\sigma$ and satisfy Dykhne's ansatz. A comparison with known numerical, asymptotic and exact results for the EC problem of an isotropic 2D regular composite showed full agreement. The obtained bounds are stronger than the known bounds in \cite{wie12,has62,nes91,che12,
che24} (the last is for a three-phase composite with one superconducting constituent).

Despite the complete correspondence between the developed theory of self-dual polynomials and different results obtained by other methods (analytical, numerical, probabilistic, variational {\em etc}) in the EC problem for the 2D three- and four-phase symmetric composites,  a number of questions remain to be answered. 
Some of these are purely mathematical in origin and were formulated in \cite{fel22}, e.g., an extension of the theory to self-dual multivariate polynomials that are invariant under the action of the finite group of color permutation $P_n\subset S_n$, such as cyclic 
${\sf Z}_n$, alternating ${\sf A}_n$ or groups' product $S_k\times S_{n-k}$.

On the other hand, some technical issues are important for the final verification, e.g., to clarify a relationship between the symmetric bounds $\Omega(\sigma_1,\sigma_2,\sigma_3)$, $\omega(
\sigma_1,\sigma_2,\sigma_3)$ and the Cherkaev bounds $\sigma_{Ch}^{\pm}(\sigma_1,\sigma_2,\sigma_3)$ in the general case of three non-superconducting phases \cite{che09}.

Finally, one more issue of the EC problem in the 2D $n$-phase composite should be mentioned. Often, we do not know the detailed description of the piecewise function $\sigma(\bf r)$ in the plane, but only the number of homogeneous phases and their volume fractions $p_j$. It is for this reason that all bounds for the isotropic effective conductivity $\sigma_e$ were derived in terms of partial isotropic conductivities $\sigma_j$ and $p_j$, which do not allow for the permutation of the variables. However, the theory of self-dual polynomials, built upon invariants of symmetric group $S_n$, can also be applied in this setting with minor modifications.

Indeed, consider the 2D $n$-phase composite with partial isotropic conductivities $\sigma_j$ and their volume fractions $p_j$, given by rational numbers $p_j=k_j/N$, $\;\sum_{j=1}^nk_j=N$. If the composite has an isotropic effective conductivity, then we can construct a commutative monoid of self-dual polynomials ${\sf R}\left(^{\lambda,\;{\bf x^N}}_{m,\;S_N}\right)$ and  ${\sf S}\left(^
{\lambda,\;{\bf x^N}}_{m,\;S_N}\right)$ in $\lambda$, of degree $mN$, where ${\bf x^N}=\{x_1,\ldots,x_N\}\in{\mathbb E}^N$. According to (\ref{e17}), a unique positive root $\lambda_{\odot}\left({\bf x}^N\right)$ of the proper self-dual polynomial ${\sf S}_{\odot}\left(^{\lambda,\;{\bf x^N}}_{m,\;S_N}\right)$ is {\em weakly} bounded by the arithmetic and harmonic means of variables 
$(x_1,\ldots,x_N)$, i.e., Wiener's bounds. If, in addition, we can find the {\em strong} bounds $\Omega(x_1,\ldots,x_N)$ and $\omega
(x_1,\ldots,x_N)$ with successive equating of partial conductivities: 
$$
\sigma_1=x_1=\ldots=x_{k_1},\qquad 
\sigma_2=x_{k_1+1}=\ldots=x_{k_1+k_2},\qquad
\sigma_n=x_{N-k_n+1}=\ldots=x_N,
$$
we would achieve the stated goal and obtain the bounds $\Omega(
\sigma_1,p_1;\ldots;\sigma_n,p_n)$ and $\omega(\sigma_1,p_1;\ldots;\sigma_n,p_n)$ for the isotropic effective conductivity $\sigma_e(
\sigma_1,p_1;\ldots;\sigma_n,p_n)$ of the 2D $n$-phase composite built of homogeneous constituents with partial isotropic conductivities $\sigma_j$ and their volume fractions $p_j$. 

\noindent
The implementation of this program is the subject of further research.
%%%%%%%%%%%%%%%%%%%%%%%%%%%%%%%%%%%%%%%%%%%%%%%%%
%%%%%%%%%%%%%%%%%%%%%%%%%%%%%%%%%%%%%%%%%%%
\section*{Acknowledgements}
I am thankful to V. Nesi, R.V. Craster and A. Cherkaev for their
communication and bringing their papers to my attention.
%%%%%%%%%%%%%%%%%%%%%%%%%%%%%%%%%%%%%%%%%%%%%%%
%%%%%%%%%%%%%%%%%%%%%%%%%%%%%%%%%%%%%%%%%%%

\end{document}